\documentclass{aa}  

\usepackage{graphicx}
\usepackage[style=base]{caption}
\usepackage{txfonts,xcolor}
\usepackage{booktabs}
\usepackage{hyperref}
\usepackage{adjustbox}
\usepackage{lscape}
\usepackage{rotating}
\usepackage{physics}

\newcommand{\HI}{\ion{H}{i}}
\newcommand{\HeI}{\ion{He}{i}}
\newcommand{\um}{$\mu$m}
\newcommand{\brg}{Br$\gamma$}
\newcommand{\kms}{km\,s$^{-1}$}
\newcommand{\rstar}{R$_*$}
\newcommand{\macc}{$\dot{M}_{acc}$}
\newcommand{\msunyr}{{M}$_{\odot}$yr$^{-1}$}

\begin{document}

   \title{The GRAVITY  young stellar object survey }
    \subtitle{XI. Imaging the hot gas emission around the Herbig Ae star HD\,58647}

\author{GRAVITY Collaboration(\thanks{GRAVITY is developed in a
collaboration by the Max Planck Institute for Extraterrestrial Physics,
LESIA of Paris Observatory and IPAG of Université Grenoble Alpes / CNRS,
the Max Planck Institute for Astronomy, the University of Cologne, the
Centro Multidisciplinar de Astrofisica Lisbon and Porto, and the European
Southern Observatory.}): {Y.-I. Bouarour}\inst{1} \and {R. Garcia Lopez} \inst{1,2} 
\and J. Sanchez-Bermudez \inst{14,2}
\and  A. Caratti o Garatti \inst{4,2}
\and K. Perraut \inst{12} \and N. Aimar \inst{9}  \and A. Amorim \inst{13,7} 
 \and J.-P. Berger \inst{12} \and G. Bourdarot \inst{8} \and W. Brandner \inst{2} \inst{8} \and Y. Cl\'{e}net \inst{9}  \and P. T. de Zeeuw \inst{11}  \and C. Dougados  \inst{12} \and A. Drescher \inst{8}
 \and A. Eckart \inst{5, 10}  \and F. Eisenhauer \inst{8} \and M. Flock \inst{2} \and P. Garcia \inst{6,7} \and E. Gendron \inst{9} \and R. Genzel \inst{8} 
\and S. Gillessen \inst{8} \and S. Grant \inst{8} \and G. Heißel \inst{16, 9} 
\and Th. Henning \inst{2} \and L. Jocou \inst{12} \and P. Kervella \inst{9} \and L. Labadie \inst{5} \and S. Lacour \inst{9} \and V. Lapeyrère \inst{9} 
 \and J.-B. Le Bouquin \inst{12} \and P. Léna \inst{9} \and H. Linz \inst{2}  \and D. Lutz \inst{8}  \and F. Mang \inst{8} \and H. Nowacki \inst{12} 
\and T. Ott \inst{8} \and T. Paumard \inst{9} 
\and G. Perrin \inst{9} \and J. E. Pineda \inst{8} \and D. C. Ribeiro \inst{8} \and M. Sadun Bordoni\inst{8} \and J. Shangguan \inst{8} \and T. Shimizu \inst{8} \and A. Soulain \inst{12} \and C. Straubmeier \inst{5} \and E. Sturm \inst{8} \and L. Tacconi \inst{8} 
\and F. Vincent \inst{9}}

   \institute{School of Physics, University College Dublin, Belfield, Dublin 4, Ireland\\
   \email{rebeca.garcialopez@ucd.ie}
   \and Max Planck Institute for Astronomy, K\"{o}nigstuhl 17, Heidelberg, Germany, D-69117     
    \and Dept. of Space, Earth \& Environment, Chalmers University of Technology, SE-412 93 Gothenburg, Sweden
  \and INAF-Osservatorio Astronomico di Capodimonte, via Moiariello 16, 80131 Napoli, Italy
  \and  I. Physikalisches Institut, Universität zu Köln, Zülpicher Str. 77, 50937, K\"{o}ln, Germany
  \and  Faculdade de Engenharia, Universidade do Porto, Rua Dr. Roberto Frias, P-4200-465 Porto, Portugal
  \and  CENTRA, Centro de Astrofísica e Gravitação, Instituto Superior Técnico, Avenida Rovisco Pais 1, P-1049 Lisboa, Portugal
  \and  Max Planck Institute for Extraterrestrial Physics, Giessenbachstrasse, 85741 Garching bei M\"{u}nchen, Germany
  \and  LESIA, Observatoire de Paris, Université PSL, CNRS, Sorbonne Université, Université de Paris, 5 place Jules Janssen, 92195 Meudon, France
  \and  Max-Planck-Institute for Radio Astronomy, Auf dem H\"{u}gel 69, 53121 Bonn, Germany
  \and  Leiden University, 2311 EZ  Leiden, The Netherlands
  \and  Univ. Grenoble Alpes, CNRS, IPAG, F-38000 Grenoble, France
  \and  Universidade de Lisboa - Faculdade de Ciências, Campo Grande, P-1749-016 Lisboa, Portugal
  \and  Instituto de Astronom\'{i}a, Universidad Nacional Aut\'{o}noma de M\'{e}xico, Apdo. Postal 70264, Ciudad de M\'{e}xico, 04510, M\'{e}xico 
    \and  Leiden Observatory, Leiden University, PO Box 9513, 2300 RA Leiden, The Netherlands
    \and Advanced Concepts Team, European Space Agency, TEC-SF, ESTEC, Keplerlaan 1, 2201 AZ Noordwijk, The Netherlands
}

   \date{Received ; accepted }
\titlerunning{HD\,58647}
\authorrunning{Bouarour, Y.-I. et al.}

 
  \abstract
   {}
   %
   {We aim to investigate the origin of the HI \brg\ emission in young stars by using GRAVITY to image the innermost region of circumstellar disks, where important physical processes such as accretion and winds occur. With high spectral and angular resolution, we  focus on studying the continuum and the HI \brg-emitting area of the Herbig star HD58647.}
   %
   {Using VLTI-GRAVITY, we conducted observations of HD58647 with both high spectral and high angular resolution. Thanks to the extensive $uv$ coverage, we were able to obtain detailed images of the circumstellar environment at a sub-au scale, specifically capturing the continuum and the \brg-emitting region. Through the analysis of velocity-dispersed images and photocentre shifts, we were able to investigate the kinematics of the HI \brg-emitting region.}
   {The recovered continuum images show extended emission where the disk major axis is oriented along a position angle of 14\degr. The size of the continuum emission at 5-sigma levels is $\sim$ 1.5 times more extended than the sizes reported from geometrical fitting (3.69 mas $\pm$ 0.02 mas). This result supports the existence of dust particles close to the stellar surface, screened from the stellar radiation by an optically thick gaseous disk. Moreover, for the first time with GRAVITY, the hot gas component of HD58647 traced by the \brg \,has been imaged. This allowed us to constrain the size of the \brg -emitting region and study the kinematics of the hot gas; we find its velocity field to be roughly consistent with gas that obeys Keplerian motion. The velocity-dispersed images show that the size of the hot gas emission is from a more compact region than the continuum (2.3 mas $\pm$ 0.2 mas). Finally, the line phases show that  the emission is not entirely consistent with Keplerian rotation, hinting at a more complex structure in the hot gaseous disk. }
   {}

   \maketitle
%

\section{Introduction}

{

The study of protoplanetary disks is of crucial importance for our understanding of the initial phases of the formation and evolution of planetary systems. In the last few years, new instruments with higher spatial resolution and sensitivity have revealed the complex structure of protoplanetary disks, showing the presence of spirals, rings,  gaps, and shadows from a few tens of au up to $\sim$200\,au from the central source \citep{Beuzit2008,Beuzit2019,HLTauALMA2015,Andrews2018,Long2018,Benisty2015, deBoer2016,Pohl2017,Benisty2017,Avenhaus2018}. These structures are thought  to arise from different phenomena, such as the presence of planets, winds, and/or disk instabilities \citep[see e.g. the recent review by][]{Benisty2022arXiv220309991B}. Most of them have been detected as close as a few au from the central source. However, it is not yet clear whether  they extend down to the innermost disk region, within $\sim$1--2\,au of the source. Some indirect evidence of the presence of structures in the innermost disk is, however, given by the so-called shadows. These regions are dark areas observed in spectro-polarimetric images obtained with extreme adaptive optics instruments such as SPHERE. They are interpreted as shadows due to misalignments between the inner and outer disk and/or the presence of scale-height variations within the disk, such as warps \citep{Pinilla2018,Benisty2018,Bohn2022}. The presence of these shadows thus allows us to place strong, albeit indirect, constraints on the complex structure of the inner disk.

Obtaining direct evidence of the inner disk structure is, however, very difficult. This is due to the small spatial scales involved, meaning optical/IR interferometry is the only technique able to spatially resolve the innermost disk. Despite the complexity of this technique, the advent of new interferometric instruments such as VLTI-GRAVITY and VLTI-MATISSE, which have higher sensitivity and better $uv$ coverage than first generation interferometers, is allowing us now to fill the gap. For example, recent spectro-interferometric surveys have revealed the presence of asymmetric $H$- and $K$-band continuum emission, which can be interpreted as unevenly illuminated rims \citep{Lazareff2017,Perraut2019}. Moreover, recent VLTI-GRAVITY and VLTI-MATISSE observations hint at the presence of rotating structures within a few au of the central source \citep{Varga2021,Joel2021}. Whereas geometrical modelling can be used to constrain the general structure of the disk, it cannot provide details on the nature of such asymmetries because geometrical models are too simple to catch the complexity of such structures.
Indeed, recent reconstructed images at near-IR wavelengths have shown different complex morphologies in the inner disk, demonstrating that disks at sub-au spatial scales are not symmetric at all and that substructures are often present \citep{Labdon2019, Kluska2020, Joel2021, Hofmann2022}.
In this context, the task becomes even more complex and/or difficult if one considers the contribution from the gas component. Indeed, within 1\,au of the source, a dust-free disk extends towards the star, and matter accretes onto the source and/or is blown away in the form of winds. Therefore, a complete picture of both gas and dust emission is only possible through direct imaging as the overall picture is too complicated to be reproduced by simple centro-symmetric geometrical models. 

So far, only one image in line emission has been reconstructed using near-IR interferometry for a young stellar object (YSO), namely the HI\,Br$\gamma$ line imaged around the Herbig Be star MWC\,297 \citep{Hone2017}. In this case, the authors find that the HI\,Br$\gamma$ line emission in MWC\,297 is consistent with having originated at the base of a disk wind, showing for the first time that kinematic effects in the sub-au inner regions of a protoplanetary disk can be directly imaged. It should also be noted that this is the only young star for which the HI\,Br$\gamma$ line emission is more extended than the $K$-band continuum emission, and therefore it is yet to be determined whether the same results can be applied to other sources with more compact \brg\ emission \cite[e.g.][]{Alessio2015, Rebeca2015, Kurosawa2016}. Sources of HI\,Br$\gamma$ line emission other than (or complementary to) disk winds are, for instance, the accreting gas and/or the disk itself.

In this paper we present the first results of an interferometric campaign carried out to obtain images of a sample of intermediate- to high-mass stars obtained with the VLTI-GRAVITY instrument as part of the YSO guaranteed time observation (GTO) programme. Here, we focus on imaging the K-band continuum and HI\,Br$\gamma$ line emission around the Herbig star HD\,58647. This object is a bright Herbig B9 IV \citep{Mora2001} star at a distance of 302\,pc \citep{GaiaEDR32021, GAIA-DR1-2016} with a luminosity of about 275\,L$_{\odot}$ \citep{Vioque2018}. Strong hydrogen emission lines have been detected and studied in detail by \cite{Manoj2002}, \cite{Brittain2007}, and \cite{Harrington2009}. One of the distinguishing features of these lines is their double-peaked profile, which hints at gas in Keplerian rotation. 

Previous optical interferometric studies of the circumstellar disk of HD\,58647 in the $H$ and $K$ bands \citep{Lazareff2017,Perraut2019} indicated the presence of an elongated continuum emission with a size of $\sim$3--4 milliarcseconds (mas), an inclination  of $\sim$65$^{\circ}$, and a position angle (PA) of $\sim$15$^{\circ}$. Interestingly, VLTI-AMBER spectro-interferometric observations successfully reproduced the double-peaked \brg\ line profile as well as the interferometric observables via a disk-wind model \citep{Kurosawa2016}. However, the lack of sensitivity and the limited $uv$ coverage of VLTI-AMBER interferometric observations did not allow for an image reconstruction of the emission line.

This paper is a step further in studying the innermost region of  HD\,58647.\ We do so by simultaneously  reconstructing an image of the $K$-band dust continuum and of HI\,\brg\ line emission using the European Southern Observatory(ESO)-VLTI beam combiner GRAVITY. In particular, we use our data to spatially and spectrally resolve both the dust and the hot gas component in the inner disk region within 1\,au of the young star.

The paper is organised as follows: Sect.~\ref{sec2} and Sect.~\ref{sec3} present our GRAVITY GTO observations and data reduction. Section~\ref{sec4} and Sect.~\ref{sec5} show  the analysis and results of the continuum and the \brg \ line emission. In Sect.~\ref{sec6} we present our discussion and in Sect.~\ref{sec7} a summary of our findings.

\begin{table*}[h]
\centering
\begin{tabular}{l c c c c c c c}
\toprule
\toprule
d [pc]                  & T$_{\mathrm{eff}}$ [K]  &  L [L$_{\odot}$]           & M [M$_{\odot}$] & Age [Ma]                & A$_{\mathrm{v}}$[mag] &  $v\sin  i$  [km/s] & \rstar [R$_{\odot}$] \\
\midrule
302.21$^{+2.43}_{-2.3}$$^{(a)}$ &10500$^{+200}_{-200}$  & 247.97$^{+63.86}_{-52.47}$                      &       3.87$^{+0.33}_{-0.19}$      &  0.84$^{+0.12}_{-0.18}$                     & 0.37$^{+0.19}_{-0.12}$ & 118$^{+4}_{-4}$$^{(b)}$ & 4.77$^{+2.03}_{-1.66}$  \\
\bottomrule

\end{tabular}
\caption{Stellar properties from \citep{Vioque2018}; the stellar luminosity is re-scaled to the \textit{Gaia} EDR 3 distance \citep{GaiaEDR32021}.}
\label{stellarpar}

\tablefoot{\noindent $^{(a)}$ \citet{GaiaEDR32021}; $^{(b)}$ \citet{Montesinos2009}}
\end{table*}

\section{Observations and data reduction}
\label{sec2}

HD\,58647 was observed with the ESO-VLTI instrument GRAVITY  \citep{Eisenhauer2011, GRAVITY2017}  as part of the YSO GTO. A full log of the observations can be found in Table\,\ref{LogObs}. Our observations were performed in six different runs, between January 2020 and February 2021. We used

the four 1.8\,m auxiliary telescopes, resulting in six baselines for each of the three adopted array configurations for a range of baseline sizes of $\sim$ 11 to 132 m. The data were recorded using both the fringe tracker (FT) and the science channel (SC) detectors covering the $K$-band range from 1.9\,\um\ to 2.4\,\um. 
The FT data were recorded at low spectral resolution (R$\sim$ 20) at a typical frame rate of 1\,kHz \citep{Lacour2019}, allowing the atmosphere effects to be frozen. The SC data were recorded at high spectral resolution (HR; R$\sim$4000) using an average DIT of 30\,s for a rough total exposure time of 300\,s per frame, and each dataset has from 7 to 19 (N in Table \ref{LogObs}) frames (see Table\,\ref{LogObs} for more details).
The complete u-v coverage of our observations is shown in Fig.\,\ref{uv-plane}, which displays sufficient sampling to attempt image reconstruction.

The data were reduced using the GRAVITY data reduction software v.1.3.0 \citep{Lapeyrere2014}. The transfer function was estimated using a calibrator star (see Table\,\ref{LogObs} for a full list of calibrators). The spectrum of the calibrator was also used to correct for the telluric absorption features present in the spectrum.  
The wavelength calibration of our data was refined by using the many telluric features present in the source spectrum before applying the telluric correction. 

{Despite the  time gap between the observations, analysis of the dataset revealed no discernible temporal variability, as depicted in Fig. \ref{fig:OBSALL}. This allowed  us to  combine the datasets, which resulted in a more robust and comprehensive analysis.}

\begin{figure}[!ht]
\centering
\includegraphics[width=\columnwidth]{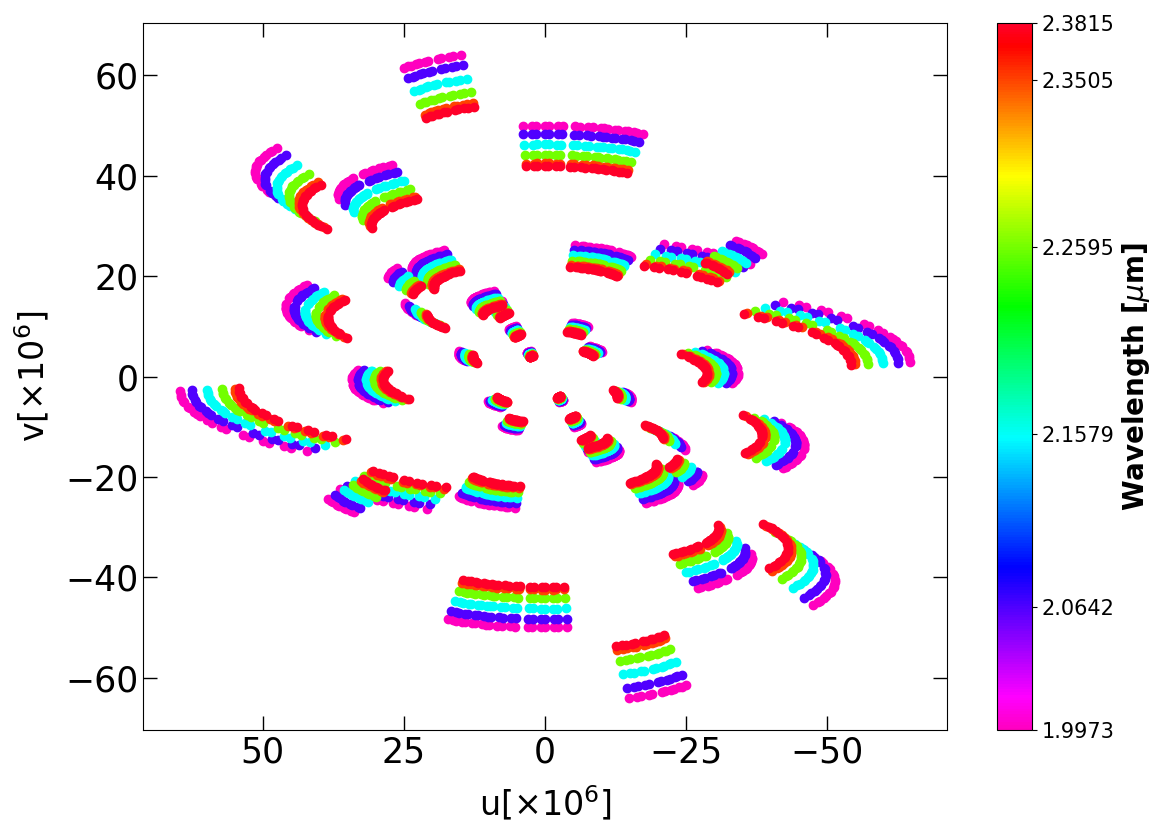}
\caption{$uv$ plane coverage of our observations of HD\,58647. See Table\,\ref{LogObs} for a detailed description of the observations.}
\label{uv-plane}
\end{figure}

\begin{figure}[!ht]

\centering
\includegraphics[width=\columnwidth]{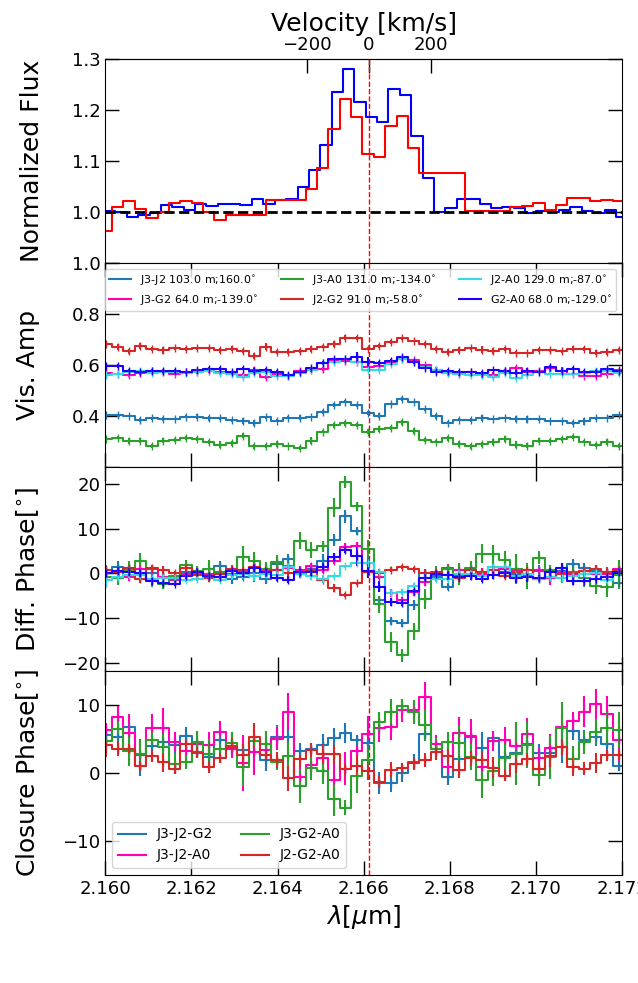}
\caption{Example of HR GRAVITY interferometric observations of HD\,58647 around the \HI\,\brg\ line position taken in January 2020 using the (J3-J2-G2-A0) configuration. From top to bottom: HR spectrum (red) and photospheric-corrected spectrum (blue); spectrally dispersed visibility amplitudes; differential phases; and closure phases. Different colours represent different projected baselines and baseline orientations (PAs; visibilities and differential phase panels) or triplets (closure phase panel), as indicated in the middle-top and bottom panel, respectively.}   

\label{diffobs}
\end{figure}

\section{Interferometric observables}

\label{sec3}
Each of our interferometric observations provides us with the following observables: spectrum, six spectrally dispersed visibilities, four closure phases, and six differential phases. Visibilities (V) measure the size of the emitting region, with V=1 and V=0 indicating spatially unresolved, or fully resolved emission, respectively. The differential phase measures the photocentre shift of the line with respect to the continuum, whereas closure phases give a measure of the departure from centro-symmetry of the emission.
An example of one GRAVITY dataset around the position of the \HI\,\brg\ lines is shown in Fig.\,\ref{diffobs}. The full sample can be found in Fig.\,\ref{fig:OBSALL}.

The spectrum of HD\,58647 shows bright \brg\ line emission (red spectrum of Top panel of Fig.\,\ref{diffobs}). Interestingly, the \brg\ line profile is double peaked with the central absorption roughly centred at zero velocity. The blue- and redshifted maxima peak roughly at $\pm$70\,\kms. In order to account for the full \HI\,\brg\ emission, the contribution from the intrinsic \brg\ photospheric feature was removed by using a photospheric template spectrum of the same effective temperature, surface gravity and  $v\sin  i$ as HD\,58647. A full description of the procedure can be found in \cite{GarciaLopez2006}. The stellar parameters and $v\sin  i$ values were taken from \cite{Vioque2018} and \cite{Montesinos2009}, respectively (see Table \ref{stellarpar}). The corrected $\mathrm{Br}\gamma$ spectrum is shown in the top panel of Fig.\,\ref{diffobs} (blue spectrum).

In the middle-top panel of Fig. \,\ref{diffobs}, we observe an overall increase in the  visibilities across the  \brg\ line. This suggests that the region emitting the  \brg\ line is more compact compared to the adjacent K-band continuum emission. Notably, the visibilities across the line are spectrally resolved and they  exhibit a double-peaked shape, with a dip at lower velocities( v$\sim$ 0 \kms), as marked by the red-dashed line.

The differential phases (Fig.\,\ref{diffobs}, middle-bottom panel) show the characteristic S-shaped profile across the \brg\ line, suggesting the presence of rotating gas. The S-shaped profile is observed in all of our baselines down to a projected baseline of $\sim$10\,m. Finally, a closure-phase signature of up to $\sim$10$^{\circ}$ across the line is observed (Fig.\,\ref{diffobs}, bottom panel). The same S-shaped profile is observed in the closure phase at the triangle configurations with the largest baseline lengths indicating departure from centro-asymmetry.

\section{Imaging the K-band continuum emission around HD\,58647}
\label{sec4}

\begin{figure*}
    \includegraphics[width=\linewidth]{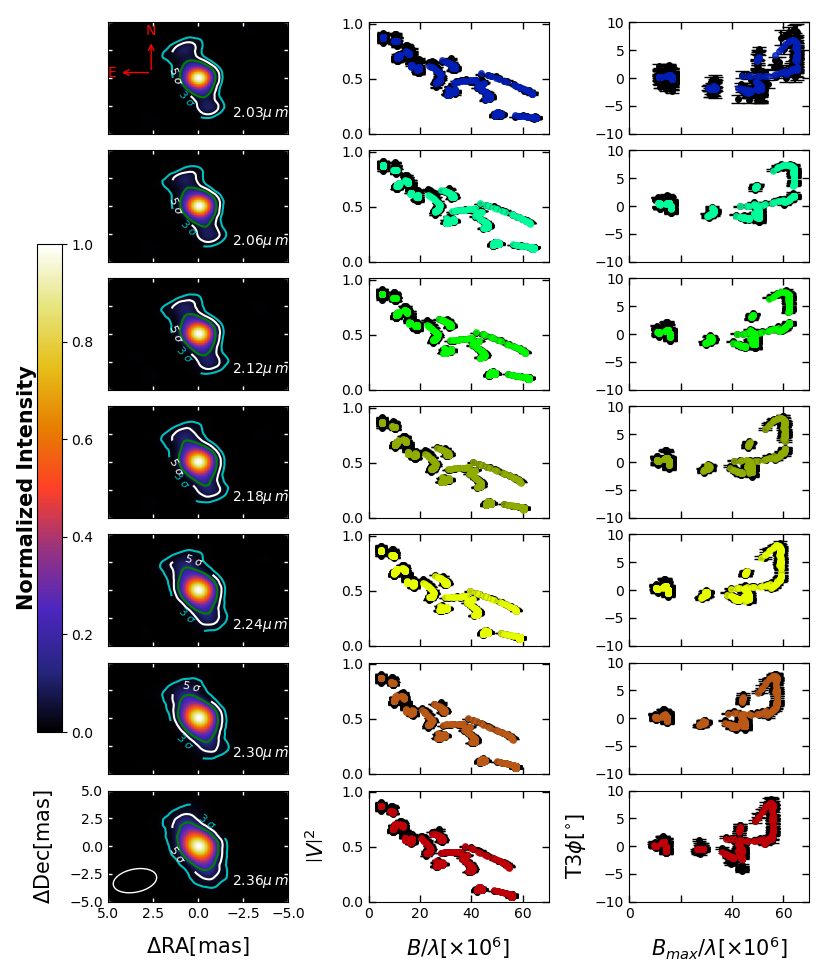}
    
    \caption{Results of the image reconstruction. {\bf Left:} K-band continuum reconstructed images of HD\,58647 at different wavelengths (reported in the labels). North is up, and east is to the left. The white hollow ellipse at the lower-left corner of the bottom panel represents the size of the clean beam. Contours represent 3, 5, and 10\,$\sigma$ pixel significance levels. {\bf Middle and right:} Squared visibilities ($\left|V\right|^2$; middle panel) and closure phases (T3$_{\phi}$; right panel)  as a function of the spatial frequency. The observed data with their corresponding error bars are represented as black dots, while the synthetic observables extracted from the reconstructed images are shown in colour.}
    \label{contIM}
\end{figure*}

\begin{table}[t]
\caption{Results from the K-band continuum image reconstruction.}
\centering
\vspace{0.1cm}
\begin{tabular}{l l c }
\hline
\hline
& R & Flux  \\
 & (mas; au) & (\%)  \\
\hline
3$\sigma$ & 3.5$\pm$0.2; 1.1$\pm$0.1 & 89  \\
5$\sigma$ & 3.2$\pm$0.2; 1.0$\pm$0.1 & 85  \\
10$\sigma$ & 1.7$\pm$0.2; 0.5$\pm$0.1 & 73  \\

\hline
\end{tabular}
\label{tab:results_cont}
\tablefoot{Column 1: contour level as in Fig.\,\ref{contIM}; Column 2: half width at half maximum of a Gaussian fit to the contour level in mas and au assuming a distance of 302.21\,pc to the source; Column 3: percentage of flux enclosed within the respective contour.}
\end{table}

Interferometric imaging is a powerful tool for obtaining a model-independent characterisation of the brightness distribution of an object in the sky. However, interferometric imaging is a challenging task as this is an ill-posed problem due to the sparsity of the u-v coverage and the non-linearity between the measured squared visibilities and closure phases. A full description of optical interferometric imaging can be found in \cite{Thiebaut2017} and \cite{Joel2018Ex}. In short, image reconstruction can be thought as minimising the following expression:

\begin{equation}
     \Vec{\mathrm{x}} = argmin\left\{ \frac{1}{2} \chi^2 \left( \pmb x\right) + \sum_i^n \mu_i R\left(\pmb x\right)_i  \right\}
    \label{ml}
,\end{equation}
where the first term is the distance to the data (also known as the negative log-likelihood),  the second term is the prior term, and $\mu$ is an hyper-parameter expressing the trade-off between the two previous terms. The solution is the sought image that best reproduces our data.

In this paper, the image reconstruction algorithm SQUEEZE \citep{Baron2010} is used to solve Eq.\,\ref{ml}. Using stochastic methods, the sought image is calculated as a combination of a large number of flux elements, ranging from 1000 to 10000, that move randomly within the grid iteratively via simulated annealing algorithms and changing the value of the criterion function in Eq.\,\ref{ml} during the exploration of the posterior distribution. The evolution of the solution as the elements move forms a Markov chain. After the stabilisation of the chain, the posterior distribution is still being explored around the best solution. This allows a mean image  to be calculated from the different chains and a standard deviation map that quantifies the significance of each pixel in our mean image to be produced. Our image was reconstructed from the HR SC data binned from the original 1634 to 7 spectral channels in order to increase the  S/N of our data.

The resolution of the images is $\lambda /2B$, with B our maximum baseline. To avoid the pixellation of the reconstructed images, we set the pixel scale to 0.15\,mas.

To initialise the different chains (initial length of 500 iterations), a synthetic image was created from a geometrical model fitting of all our interferometric observables. The geometrical model was also based on the SC HR data binned to seven spectral channels. A full description of the geometrical model is shown in Appendix\,\ref{sect:geometrical_modelling}. 

The final mean images and error maps were created from the results of all the converged chains. We used a combination of two regularizers, l$_0$ norm and total variation (TV), for all the images in the continuum.

The reconstructed images are presented in the left panel of Fig.\,\ref{contIM} with 3, 5, and 10\,$\sigma$ contours. As shown in Fig.\,\ref{contIM} (right panel),  the presented images reproduce our observables well.

The resulting image shows an elongated emission in the direction perpendicular to the elongated beam and it strongly  suggests the presence of a thin disk that is oriented in the NE--SW plane with size at 3$\sigma$ of  R$\sim$3.5\,mas (i.e. $\sim$1.1\,au), inclination of i$\sim$65\degr\ and  PA from north to east $\sim$15\degr,  (Table\,\ref{tab:results_cont}). The size of the extended emission is well beyond our element of resolution (i.e. the clean beam, white ellipse at the left-bottom of Fig.\,\ref{contIM}). The clean beam is extracted from the dirty beam by Fourier transform of the u-v plane. The clean beam size is 2.6\,mas$\times$1.9\,mas with a PA$\sim$122$\degr$.

The continuum emission is roughly symmetric within 10\,$\sigma$ ($\sim$1.5$\times$0.9\,mas from the source, i.e. 1.1$\times$0.5\,au at the distance of HD\,58647), although a slight increase in the elongation with wavelength is observed in the direction perpendicular to our element of resolution. However, the extended emission is clearly asymmetric when exploring larger spatial scales (e.g. 5\,$\sigma$ to 3\,$\sigma$ contours). 
These asymmetries cannot be reproduced by simple geometrical modelling, as the one presented in  Appendix\,\ref{sect:geometrical_modelling} namely the models are not able to perfectly reproduce the interferometric observables, in particular, the closure phase signatures (see Fig.\,\ref{v2cp}).

\section{Imaging the HI\,$\mathrm{Br}\gamma$ line emission in HD\,58647}
\label{sec5}

\begin{figure*}[th!]
    
    \includegraphics[width=\linewidth]{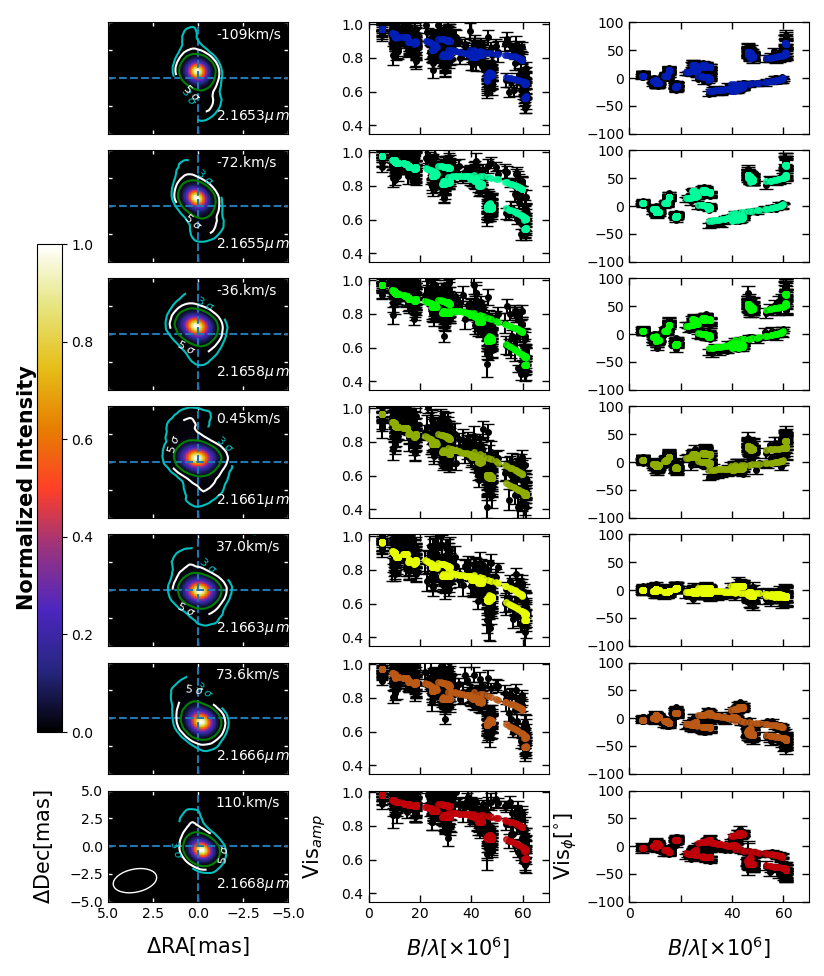}

    \caption{Results of the image reconstruction. {\bf Left:} \HI\,\brg\ line reconstructed images of HD\,58647. North is up, and east is to the left. Contours represent 3, 5, and 10 $\sigma$. {\bf Middle and right:} Comparison of the observed  continuum-subtracted visibility amplitudes and absolute  phases (black circles with error bars) with the synthetic observables extracted from the reconstructed image.}
    \label{velomaps}
\end{figure*}

Imaging across IR line tracers is even more challenging than obtaining images of the continuum due to the reduced number of spectral channels available to obtain an image. Therefore, very few images across the \brg \, line are available in the literature, and the very few available usually involve very bright objects. Of the few examples, imaging across the \HI\,\brg\ and Br$\alpha$, and \HeI\ lines has been possible in Eta\,Car \citep{JSB2018, Weigelt2021}, and across the \HI\,\brg\ line in the Herbig B[e] star MWC\,297 \citep{Hone2017} the only YSO, so far.

Most of these images were reconstructed using the {IRBIS} method  \citep[see][]{Hofmann2014,Hofmann2016}, in combination with the differential-phase method detailed in \cite{Millour2011}, \cite{Weigelt2016}  in order to obtain velocity-dispersed images of the total emission (line emitting gas plus continuum) across the line. In some cases, the continuum contribution is later subtracted from the resulting image  to increase the contrast of the line emitting region and to better study the gas morphology.

In the case of HD\,58647, we opted to adopt a slightly different approach to recover the  brightness distribution across the \brg\ line. Instead of removing the continuum contribution a posteriori, after the image is reconstructed, we reconstructed the velocity-dispersed image from continuum-subtracted visibilities and differential phases. It should be noted that, even if continuum-subtracted, the displacements obtained from the observed differential phases are with respect to the continuum, assumed to be centro-symmetric. Therefore,  to take into account possible asymmetries in the brightness distribution of the continuum and line emission, the \brg\ line velocity-dispersed image has to be reconstructed from absolute displacements, that is, the Fourier phase across the \brg\ line has to be retrieved from the data. In principle this cannot be done, as optical interferometry does not allow us to obtain a one-to-one match between visibilities and phases. However, in our case, the Fourier phase can be retrieved making use of the continuum image presented in Fig.\,\ref{contIM}, and subtracting this contribution from the observed continuum-subtracted differential phases across the line (see Appendix \ref{sect:continuum-subtracted} in \citealt{Weigelt2007} for a detailed description of the interferometric observables).

In this way, the linear combination of the Fourier phases from the continuum images and the continuum-corrected differential phases allow us to derive the absolute phases across the \brg\ line. This along with the continuum-subtracted visibilities  is then used to compute continuum-subtracted \brg\ line velocity-dispersed images. A description on how the continuum-subtracted \brg\ line visibilities and differential phases have been retrieved can be found in Appendix\,\ref{sect:continuum-subtracted}.
 
To obtain the iso-velocity images across the \brg\ line the software SQUEEZE was used in the same fashion as for the continuum image reconstruction, including the same grid parameters and regularizers. 
The resulting images across the \brg\ line are shown in Fig.\,\ref{velomaps}.

The images were only reconstructed across channels where the flux of the  emission line is 10\% larger than  the continuum flux.
This resulted in images across seven spectral channels with velocities ranging from -109 \kms\ to 110 \kms\ . The right panel in Fig.\,\ref{velomaps} shows a comparison between the observed continuum-subtracted visibilities and absolute phases, and those extracted from the retrieved \brg\ line images. As shown in the middle and right panels of Fig.\,\ref{velomaps}, our images recover the observed quantities well within the errors. 

Our recovered velocity-dispersed images show marginally resolved \brg\ line emission at all spectral channels with an average size of the line emitting region at 5\,$\sigma$ level of $\sim$2.3\,mas$\pm$0.2\,mas (i.e. 0.7 au $\pm$ 0.06 au or   $\sim$31.3\,\rstar$\pm$2.7\,\rstar). No significant variations of the size are observed as a function of the spectral channel. 
 
However, each velocity-dispersed image shows photocentre shifts from the north-east direction towards the south-west from the blueshifted channels { (2.1653\,\um $\leq \lambda \leq$2.1658\,\um; -109\,\kms $\leq $v$ \leq$-72\,\kms)} to the redshifted channels { (2.1666\,\um $\leq \lambda \leq$2.1668\,\um; 73\,\kms $\leq$v  $\leq$110\,\kms)}, going through the centre of the image at the two central velocity channels { (2.1661\,\um $\leq \lambda \leq$2.1663\,\um; -36\,\kms $\leq$v$\leq$37\,\kms)}, suggesting gas in rotation.

\begin{figure}[th!]
\centering

\includegraphics[width=\columnwidth]{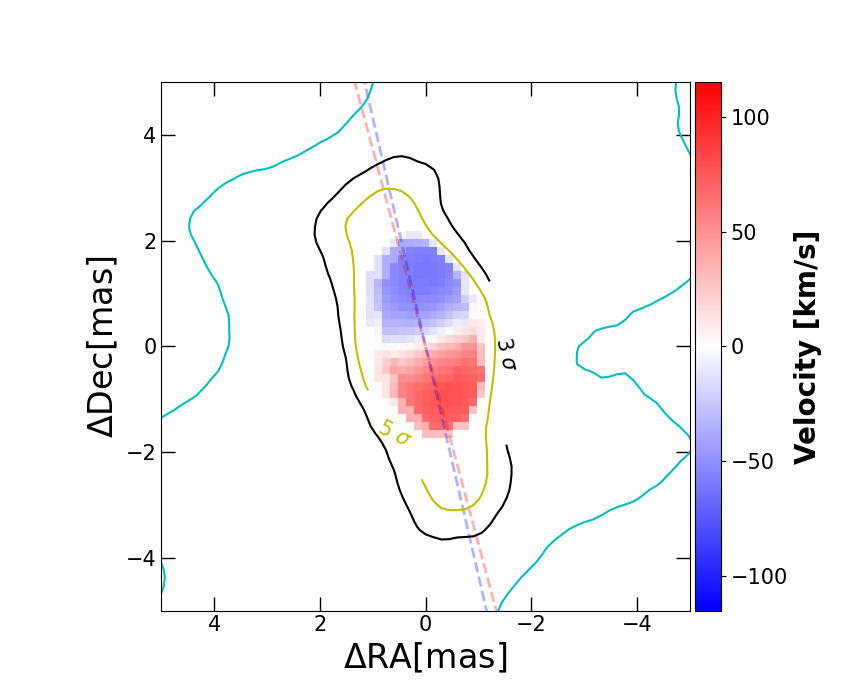}
\caption{ \brg\ line first moment map (M1)  obtained from the line velocity-dispersed images. For comparison, the 1, 3, and 5 $\sigma$ K-band continuum contours are overplotted in cyan, black, and yellow. The dashed red and blue lines illustrate the orientation of the semi-major axis of the continuum and the \brg \, line emission, respectively.}
\label{M1}
\end{figure}

To further probe the kinematic structure of the system, we computed the first moment map (M1) from the recovered velocity-dispersed \brg\ line images. In doing so, only emission detected above 5 $\sigma$ was considered. In addition, the new intensity maps were scaled to the flux of the observed GRAVITY spectrum. The result is shown in  Fig.\,\ref{M1}. The resulting map supports our findings from the \brg\ line images shown in Fig.\,\ref{velomaps}: the blueshifted emission is shifted towards the north-east, whereas the redshifted emission is shifted towards the south-west.
{Given the elliptical shape of the emission, and assuming that the \brg\ line is emitted in a disk-like structure,  inclination and PA of the \brg\ line emission can be derived. Assuming a circular structure at zero inclination, an inclination of  55\degr\ $\pm$ 1 \degr is derived. This value is very similar to the one obtained from the continuum. In the same fashion, the major axis of the disk can be estimated by measuring the location of the blueshifted and redshifted peaks and measuring the angle between them. We find a value of $\sim$13\degr, very similar to the value of $\sim$15\degr\ obtained from the K-band continuum image.} 
Furthermore, the PA obtained from the \HI\,\brg\ line velocity-dispersed images, is also in agreement with the astrometric displacements computed from the continuum-corrected differential phase signatures (see a full description of this procedure in Appendix\,\ref{sect:continuum-subtracted}). The \brg\ line astrometric displacements (see Fig.\,\ref{2ddisp}) clearly align along a straight line with PA=15\degr$\pm$5\degr, and  show again a clear shift of the blue- and redshifted emission towards the north-east and south-west, respectively.

\section{\HI\,\brg\ line Keplerian disk modelling}

The results from our \brg\ line reconstructed images, the S-shaped differential phases, and the \brg\ line astrometric displacements points towards the presence of rotating gas emitting in \HI\,\brg\ line located in the inner gaseous disk. To give more quantitative information about the hot gas kinematics, a simple \brg\ Keplerian disk model was computed and compared with our results. 

The Keplerian disk velocity field was derived assuming a M$_*$=3.87\,M$_{\odot}$ \citep{Vioque2018}. The model was created to match the angular and spectral resolution of our GRAVITY observations. The inclination, PA, and size of the disk model are set as free parameters.

Figure\,\ref{ccdiffph} shows a comparison between the observed continuum-corrected \brg\ line differential phases and those obtained from our best fitting model. 
Our best fitting model is obtained for a \brg\ line disk with an inclination and PA of $\sim$52\degr$^{+7}_{-4}$, and $\sim$14\degr$\pm$8\degr, respectively, and a radius of the emitting region of $\sim$1.1\,mas$\pm$0.3\,mas (i.e. $\sim$0.3\,au).
The PA is in agreement with that found from the \brg\ line image and K-band continuum reconstructed images, indicating no major misalignment between the continuum and line emitting region.
As seen in Figs.\,\ref{ccdiffph} and \ref{ccdiffph_old}, the model is able to reproduce the general trend observed in our data.  
However, there is a slight mismatch between the model and the observed values ($\chi^2_{\nu}\,\sim 2$). On the one hand, lower spatial frequencies are sampled better than the highest ones, hinting to a more complex structure at small spatial scales. On the other hand, the best match between observed and synthetic values is obtained for intermediate velocities (e.g. $\sim$-73\,\kms, $\sim$37\,\kms, and $\sim$74\,\kms). Finally, the redshifted velocities (but for the highest velocity channel) are roughly fitted better than blueshifted channel velocities, possibly hinting to the presence of an additional blueshifted velocity field. This might be also supported by the asymmetric double peaked \brg\ line profile showing brighter emission at blueshifted velocities.

\begin{figure}
   
    \includegraphics[width=\columnwidth]{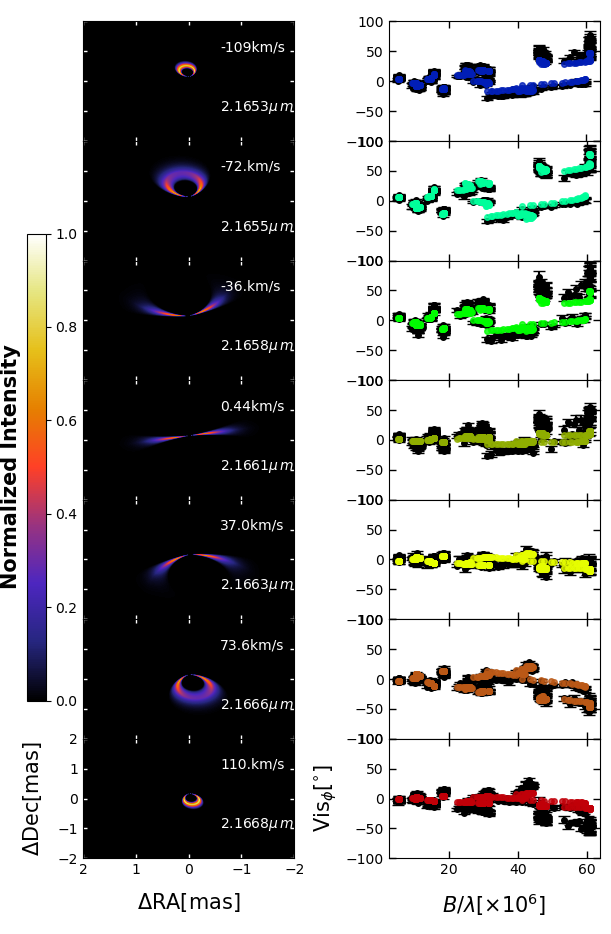}
    \caption{Results of the Keplerian disk model fitting. {\bf Left:} Velocity maps from our \brg\ line Keplerian disk model. From top to bottom, we show velocity maps at: $\sim$-110\,\kms, $\sim$-73\,\kms, $\sim$-36\,\kms, $\sim$0\,\kms, $\sim$37\,\kms, $\sim$74\,\kms, and $\sim$110\,\kms. See the main text for details. {\bf Right:} Comparison between the observed continuum-corrected \brg\ line absolute phases (black dots) and those obtained from the \brg\ line Keplerian disk model shown in \emph{left} panel (coloured dots). }
    \label{ccdiffph}
\end{figure}


\section{Discussion}
\label{sec6}

\subsection{K-band continuum emission}

Figure\,\ref{contIM} shows our continuum K-band reconstructed image (left panel) along with a comparison with our interferometric observations (middle and right panels). From this image we extracted the K-band continuum size, inclination, and PA. The inclination $\sim$64\degr\ and PA$\sim$14\degr\ derived from the image are similar to the ones found from the geometrical model presented in Appendix\,\ref{sect:geometrical_modelling} and the ones reported in \cite{Kurosawa2016,Lazareff2017, Perraut2019}. Despite the similarities between the final image and the simple 2D Gaussian model, the image is able to better recover the interferometric observables than the geometrical model (see Figs.\,\ref{contIM} and \ref{2dGau}), especially at the largest spatial scales. This might be due to the presence of extended emission in our image (the 5-sigma contour represents $\sim$90\% of the flux) that  cannot be properly accounted for using simple single-component Gaussian and/or Lorentzian brightness distributions. This, in turn, allows us to provide a better estimate of the size of the K-band continuum emission. Moreover, image reconstruction allows us to reproduce the small asymmetries in the continuum as well as to directly derive the \brg \, brightness distribution. Indeed, our recovered image shows that the size of the K-band continuum emission extends as far as R$\sim$3.2\,mas at $\sim$90\% flux level (5$\sigma$ contour). This is a factor of $\sim$1.5 more extended than the value estimated from the geometrical modelling presented in Appendix\,\ref{sect:geometrical_modelling}, and derived in previous studies  \cite[e.g.][]{Lazareff2017,Perraut2019}. At high flux contributions ($\sim$70\%, $\sim$10$\sigma$ contour), the size from the geometrical modelling and our image roughly reconcile to each other. 
Our image shows thus K-band continuum emission from R$\sim$3.2\,mas (i.e. $\sim$0.97\,au) down to our resolution limit of $\sim$1.9\,mas (i.e. $\leq$0.6\,au) as estimated from the clean beam (see Sect.\,\ref{sec4} and Fig.\,\ref{contIM}).

This range of K-band continuum emission is in agreement with hydrodynamic (HD) and magneto-hydrodynamic models of the dust silicate inner rim  \cite[see e.g.][]{Flock2016,Flock2017}. These models include a constant gas inflow that replenish the inner disk of small dust particles and gas, and they take into account the absorption of stellar radiation by the inner gaseous disk interior to the dust sublimation front, backwarming by IR radiation, as well as accretion stress and heating. Under these conditions, the rim has a triangular shape that extends over a relative high spatial extent, showing a tip that lays in the disk mid-plane and points towards the star. The extent of the inner rim triangular shape, both in height and radially, mainly depends on the stellar luminosity. This determines the average location of the dust sublimation radius  \citep[e.g. ][]{Dullemond2010, Isella2005} and the mass accretion rate, which affects the location of the innermost tip of the triangular rim (R$_{rim}^{in}$;\cite{Flock2016, Flock2017}). In general, the higher the stellar luminosity the further the location of the rim with respect to the star is. In the same way, the higher the accretion rate the closer R$_{rim}^{in}$ is located and the higher the scale-height of the rim is.
In our case, the luminosity of HD\,58647 is log\,L$_*$=2.4 and the mass accretion rate ranges between \macc$\sim$ 3.5$\times$10$^{-7}$\,\msunyr-10$^{-6}$\,\msunyr- (\citealp{Sierra2022ApJ,Brittain2007}). The combination of these values will on the one hand bring  R$_{rim}^{in}$ closer to the star, and on the other hand, move further away the location of the sublimation front (intended here as the location of the local maximum in aspect ratio of the IR disk photosphere; see R$_{rim}^{out}$ \citealt{Flock2016}). 

The measured R$_{rim}^{out}\sim$1\,au is indeed in general agreement with the location of the rim in classical rim models including an optically thin inner disk in between the star and the rim location with dust temperatures between 1000\,K and 1500\,K and with or without backwarming \citep[see e.g. Fig.\, 5 in][]{Koumpia2021}. On the other hand, the high accretion rates might explain the presence of K-band continuum emission within R$_{rim}^{out}$ and down to at least 0.6\,au. With these high accretion rates R$_{rim}^{in}$ will move very close to the stellar surface. Alternatively, if the highest available value of \macc=10$^{-6}$\,\msunyr\ is considered, the inner gaseous disk might be optically thick \citep{Muzerolle2004}. If this is the case, dust particles might be screened from the stellar radiation and survive very close to the stellar surface, or the gas itself might even be responsible for part of the K-band continuum emission.
Therefore, our images seem to support the presence of a silicate triangular shape rim as the one described in \citep{Flock2016} extending from R$_{rim}^{in}\leq$0.6\,au up to R$_{rim}^{out}\sim$1\,au.

\subsection{\HI\,\brg\ line emission}

We have obtained the first \HI\,\brg\ line image reconstruction of the hot gas component around the Herbig Ae star HD\,58647 and one of the few reconstructed images of the gas component around a YSO (Sect.\,\ref{sec5}). Our image reconstruction method retrieves the Fourier phases from the continuum image, allowing us to obtain a more accurate estimate of the photocentre shifts of the \brg\ line. It should be recalled that the photocentre shifts obtained from the \brg\ line reconstructed image are `absolute', that is, they no longer are a function of the photocentre of the line with respect to the continuum. 
As the \brg\ line is only marginally resolved, we cannot provide information about variation of the size as a function of the spectral channel, and we can only give an average size of the \brg\ line emitting region of $\sim$2\,mas, that is, $\sim$30\,\rstar. However, our images are sensitive to changes in the photocentre shift of the line with velocity. These changes are supported by the first moment map, as well as the astrometric displacements of the line with respect to the continuum photocentre (Figs. \ref{M1} and \ref{2ddisp}, respectively). These measurements clearly show a shift between the blue- and redshifted \brg\ line emission pointing out to the presence of rotating gas around HD\,58647. 

The hot gas component as traced by the \brg\ line emission would partially overlap with a fraction of the K-band continuum emission as measured in Sect.\,\ref{sec4}. This is in contrast with previous results that reported the size of the \brg\ line emitting region smaller than that of the continuum \citep{Kurosawa2016, RebecaDec2022} . It should be noted however, that, as already discussed in the previous section, our image reconstruction allows us to obtain a better estimate of the full extension of the emission than simple geometrical models do. That said, our limited baseline length does not allow us to spatially resolve the innermost disk regions and we cannot infer whether the line emission extends down to the stellar surface (as suggested by a boundary layer scenario) or is truncated by a magnetosphere. 

Taking advantage of the photocentre displacements, and the Keplerian modelling shown in Sect.\,\ref{sec6}, we see indications that the line emission is extending down to at least $\sim$1.1\,mas, or $\sim$15\,\rstar. The extension of a possible magnetosphere around HD\,58647 is expected to be smaller than the corotation radius in order to allow accretion to proceed. The corotation radius can be estimated as $R_{cor}=(G M_*)^{1/3} (P_*/2 \pi)^{2/3}$= 2.3\,\rstar, with $P_*=2\pi R_*/ v_{eq}$= 2.1\,day. On the other hand, 
\cite{Jarvinen2019} measured a weak longitudinal magnetic field in HD\,58647 of $B_z\sim$209\,G. This small magnetic field, along with the fast rotation of the star (v\,sini=114\,\kms; \cite{Montesinos2009}) would place the location of a possible magnetosphere very close to the stellar surface. Therefore, we could hypothesise that the extension of the accretion funnel could be from as close as 1.3\,\rstar--1.5\,\rstar\ \citep{Kurosawa2016, Rebeca2015, Alessio2015} up to the location of the corotation, that is, $\sim$2.3\,\rstar. Therefore, the presence of the magnetosphere could not explain the full extension of the \brg\ line emission.

\cite{Kurosawa2016} were, however, successful in reproducing VLTI-AMBER high resolution (R=12\,000) observations of HD\,58647 with a model including emission from a compact magnetosphere and disk wind. The compact magnetosphere would produce a small absorption centred at zero velocity, helping to better reproduce the total line flux at zero velocity, whereas the disk wind would contribute for most of the emission and it will extend up to  23.5\,\rstar. This value is close to our measured size of the \brg\ line emission of $\sim$30\,\rstar. 
If this is the case, the fact that the \brg\ line profile is double peaked, and that our Keplerian disk model presented in Sect.\,\ref{sec6} roughly reproduces the line differential phase signatures, indicates that the wind is emitted very close to the disk surface where the poloidal component is not yet important. Alternatively, we cannot discard that the surface layers of the disk itself could be at the origin of at least part of the \brg\ line emission. Variations in scale-height on the origin of the \brg\ line emission could also reproduce the observed asymmetric \brg\ line profile. At the same time, this could also explain the small disagreement between our Keplerian disk model and the measured line photocentre shifts, as tiny variations on the depth at which the \brg\ line is emitted will produce displacements from the Keplerian velocity as measured from the disk mid-plane  \citep[see e.g.][]{Backs2023}.  Unfortunately, due to the limited spectral resolution, our observations do not allow us to distinguish between bound and unbound gas.

On the other hand, our measured size of the \brg\ line emission seems to disagree with the presence of a photo-evaporative wind as the origin of most of the emission. According to photo-evaporative models, and taken into account the mass of the star, even for the case of extreme UV photo-evaporative wind the bulk of the wind emission would be located much further out than our measurements (i.e. R$_{EUV}>$5\,au, that is, R$_{EUV}>$200\,\rstar).

\section{Summary}
\label{sec7}

In this paper we present  GRAVITY observations of the dusty and gaseous disk around the Herbig star HD 58647. Our dataset from 2020 and 2021 includes three different telescope configurations (small, medium, and large) that cover various baseline lengths and orientations ($\sim$ 11 to 132 m). These observations have allowed us to marginally resolve the target in the $K$ band at HR. The acquired data show no variability in terms of the wavelength-dependent observables, enabling us to combine the entire dataset. Image reconstruction has proven useful in studying the morphology of the innermost region  of HD 58647, including both the continuum and HI Br-gamma emission. Based on the analysis of the recovered images, we draw the following conclusions:

\begin{itemize}
    \item The geometry of the marginally resolved  continuum $K$-band emission  at a significance level of 3$\sigma$ is similar to our  result from the geometrical model. The orientation of the major axis of the disk has a PA of $\sim$ 14\degr \ and an  inclination of 64\degr. 

    \item Although the size of the  $K$-band continuum emission at the 3$\sigma$ level is larger than that of the geometrical model, the emission enclosed within the contour represents 90\% of the total flux, which is in good agreement with the model. This model predicts that the flux fractions contributed by the unresolved star and the circumstellar disk are approximately 91\%.

    \item The measured extent of the emission from our continuum images aligns with the location predicted by classical rim models. These models assume the presence of a thin inner disk between the star and the rim, with dust temperatures ranging from 1000 K to 1500 K.

\end{itemize}

We also present a novel technique for recovering the brightness distribution across the \brg \, line and reconstructing the \brg-dispersed images. Our findings from the analysis of the velocity-dispersed images  across the \brg\ line are as follows:

\begin{itemize}
    \item There is no significant misalignment observed between the hot gaseous disk, traced by the \brg\ line emission, and the continuum disk. The orientation of the \brg-emitting region is similar to that obtained from the astrometric displacement calculated using the differential phases and the continuum images.
    
    \item The velocity field of HD58647 has been recovered thanks to the reconstructed images across the \brg\ line. The first moment map shows blueshifted emission towards the north-east, while the redshifted emission originates in the south-west. Both emissions are oriented along the major axis of the disk.

    \item The origin of the emission is still not entirely understood. However, when compared with a disk in Keplerian rotation, we find residuals; this suggests the presence of an unresolved complex structure in the hot gaseous disk of HD58647. This is compatible with a disk wind emitted close to the disk surface; in other words, we may be detecting the base of a disk wind.

\end{itemize}

\bibliographystyle{aa} 
\bibliography{references.bib}
\begin{acknowledgements}
The authors extend their gratitude to the anonymous referee for their insightful and valuable feedback.
This material is based upon works supported by Science Foundation Ireland under Grant No. 18/SIRG/5597. A.C.G. has been supported by PRIN-INAF MAIN-STREAM 2017 “Protoplanetary disks seen through the eyes of new generation instruments” and from PRIN-INAF 2019 “Spectroscopically tracing the disk dispersal evolution (STRADE)”.  J.S.-B. acknowledges the support received from the UNAM PAPIIT project IA 105023; and from the CONAHCyT “Ciencia de Frontera” project CF-2019/263975. This research has made use of the NASA Astrophysics Data System, CDS Astronomical Data bases SIMBAD and VIZIER \footnote{Available at http://cdsweb.u-strasbg.fr/} and of the Jean-Marie Mariotti Center \texttt{Aspro and SearchCal} service\footnote{Available at http://www.jmmc.fr/aspro}.\\

\end{acknowledgements}

\begin{appendix}

\section{Observation log}
The log of the observations is given in Table \, \ref{LogObs}.
\begin{table*}[th!]
\caption{HD\,58647 GRAVITY observations. UD diameters were obtained from JMMC SearchCal \citep{Bonneau2006A&A...456..789B, Bonneau2011A&A...535A..53B}}
\centering
\begin{tabular}{l c c c c c }
\toprule
\toprule
Date       & Calibrator      & Calibrator UD Diameter  [mas]&  Array  & N\\
\hline
27-01-2020 & HD 57939, HD 60325   & 0.454, 0.121& A0-G2-J2-J3& 9\\
            
\midrule
28-01-2020 & HD 44423,HD 60325  &0.256, 0.121& D0-G2-J3-K0& 7\\

\midrule
04-02-2020 & HD 65810,HD 103125   &0.434, 0.851& A0-B2-C1-D0& 12 \\

\midrule
23-12-2020 & HD 57939,HD 60325  &0.454, 0.121& D0-G2-J3-K0& 19\\
            
\midrule
10-02-2021 & HD 57939,HD 99015  &0.454, 0.292& D0-G2-J3-K0& 12\\

21-02-2021 & HD 114461,HD 60325  &0.383, 0.121& A0-B2-C1-D0& 9\\
            
\bottomrule
\end{tabular}
\label{LogObs}

\end{table*}

\section{Observational data}
Figure\, \ref{fig:OBSALL} shows the VLTI-GRAVITY observations of HD5867 around the wavelength of the \brg \, line. Each figure shows the results of merged data for  different dates, as displayed in the caption. There are three panels, showing the visibility amplitude  (left), differential phase (middle) and differential closure phase (right). The top panels show the  photospheric-corrected spectrum. The
length and PA of the projected baselines are given in the left panels.

\begin{figure*}[!th]
    \centering
    \includegraphics[width=\linewidth]{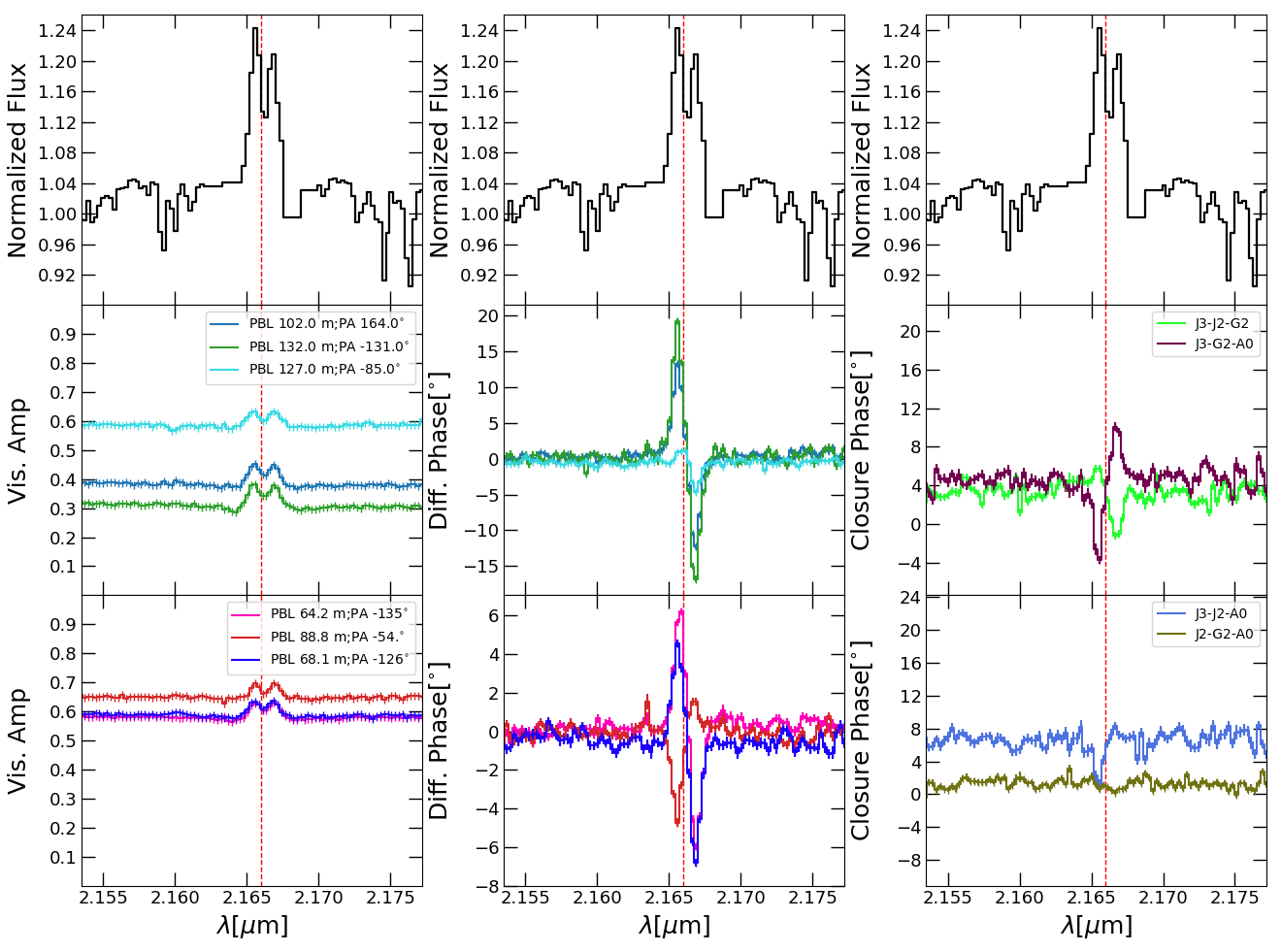}
    
    \caption{Merged spectrum, visibilities, differential phases, and closure phases of the January 27, 2020, GRAVITY interferometric data.}
    \label{fig:OBSALL}
\end{figure*}

\begin{figure*}[!h]
    \centering
    
    \includegraphics[width=\linewidth]{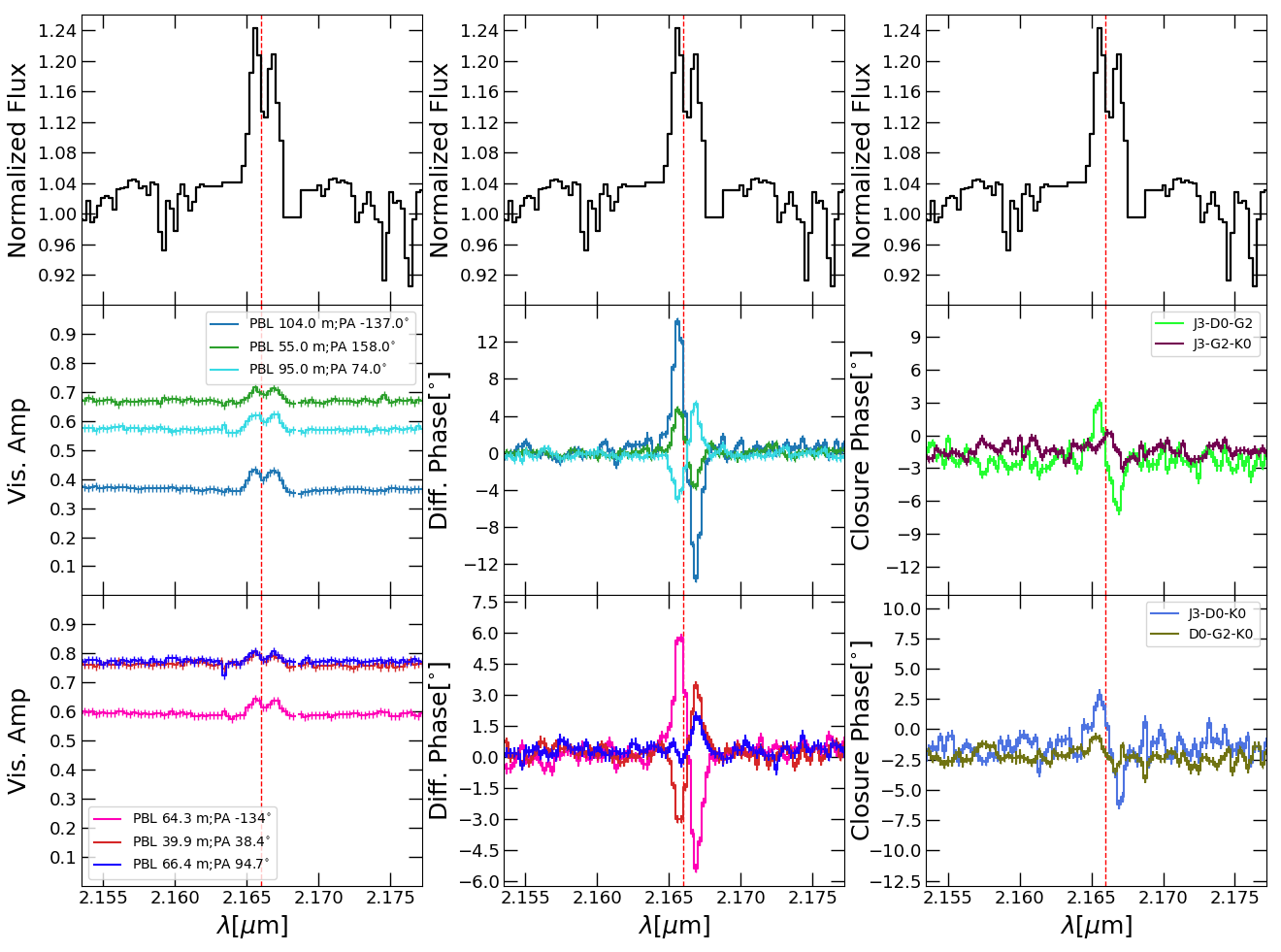}
    
    \caption{Same as Fig.\ B.1, but for January 28, 2020.}
\end{figure*}

\begin{figure*}[!h]
    \centering
    
    \includegraphics[width=\linewidth]{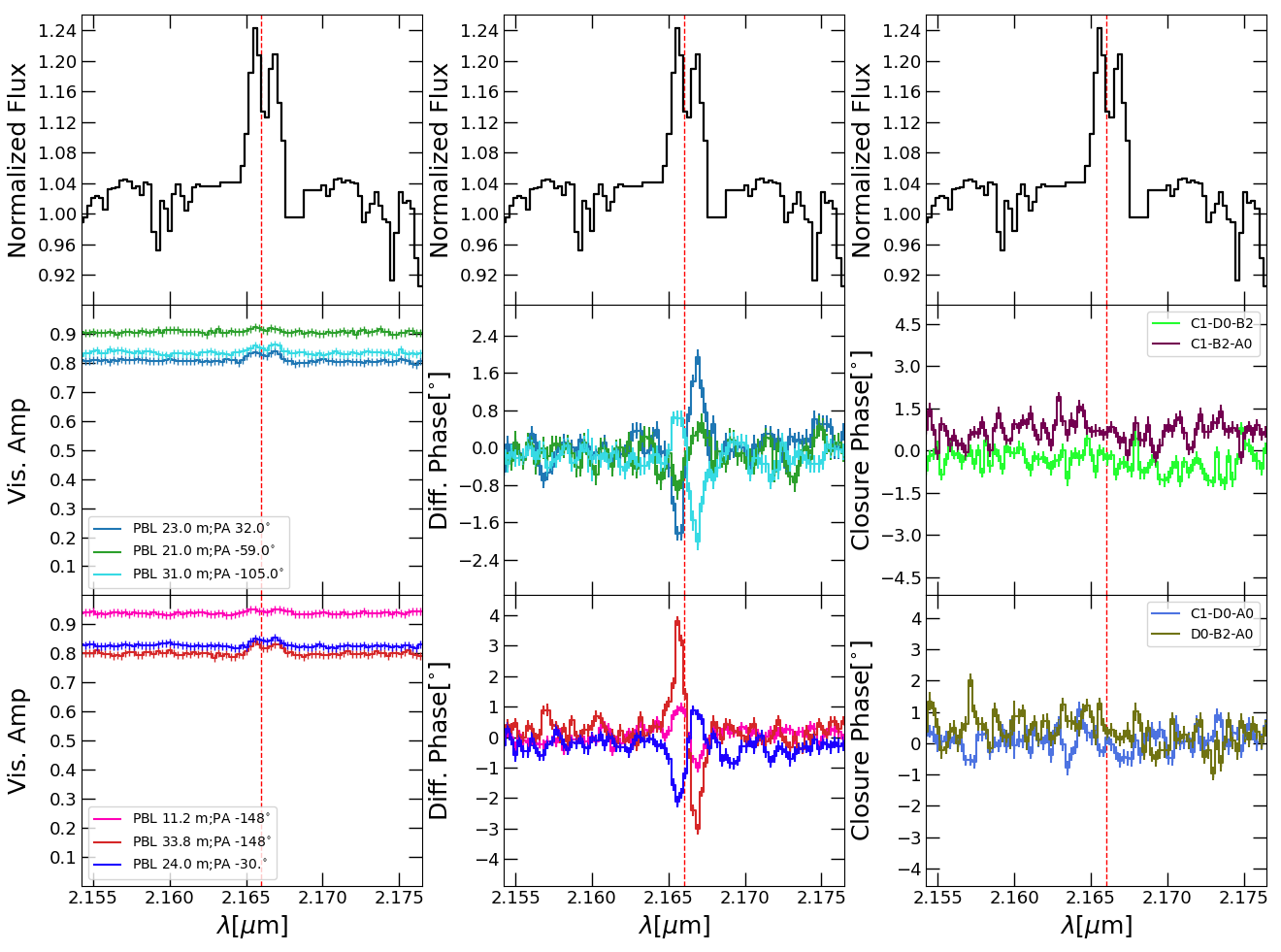}
    \caption{Same as Fig.\ B.1, but for February 4, 2020.}
\end{figure*}

\begin{figure*}[!h]
    \centering
    \includegraphics[width=\linewidth]{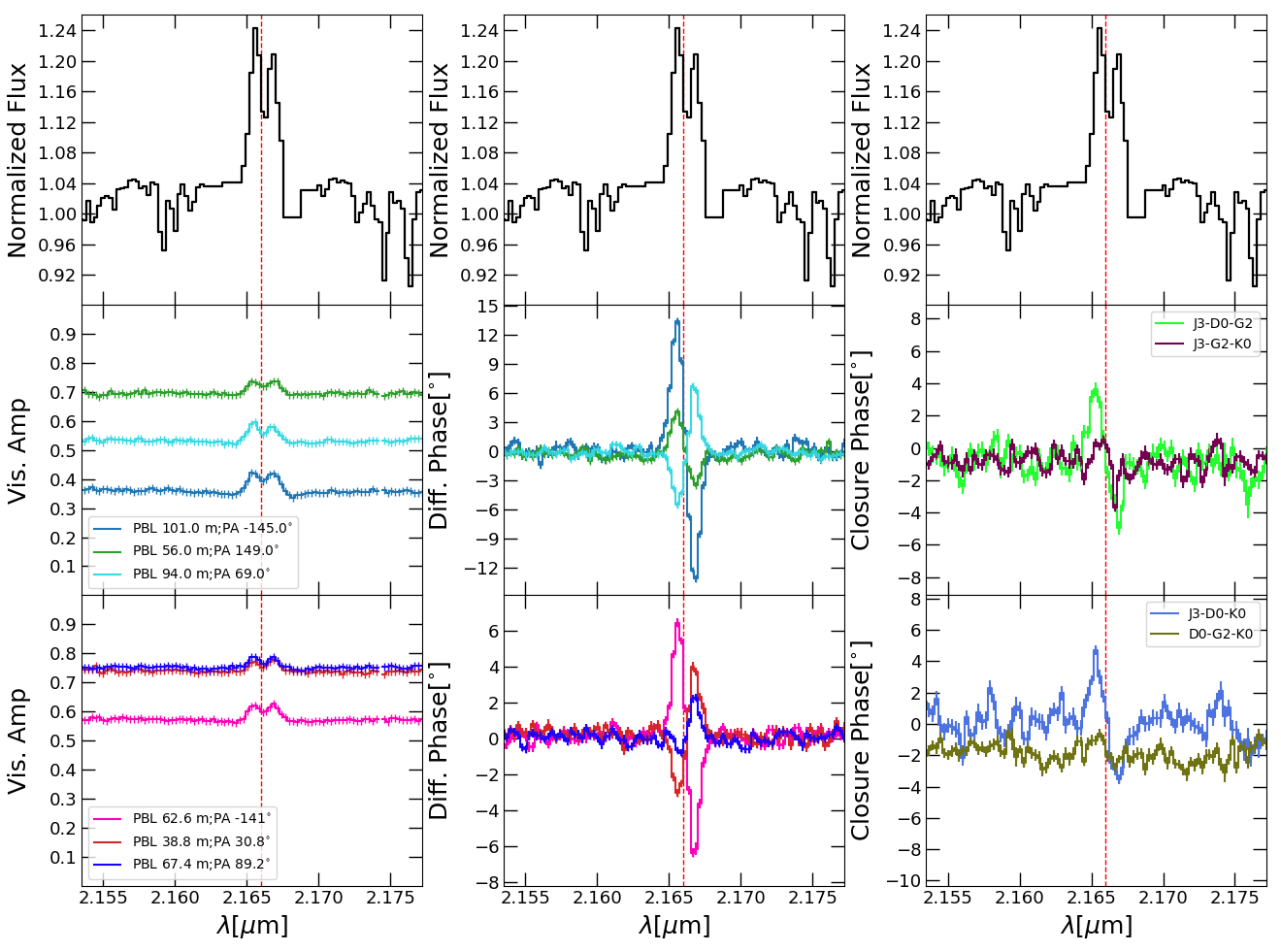}
    
    \caption{Same as Fig.\ B.1, but for December 23, 2020.}
\end{figure*}

\begin{figure*}[!h]
    \centering
    \includegraphics[width=\linewidth]{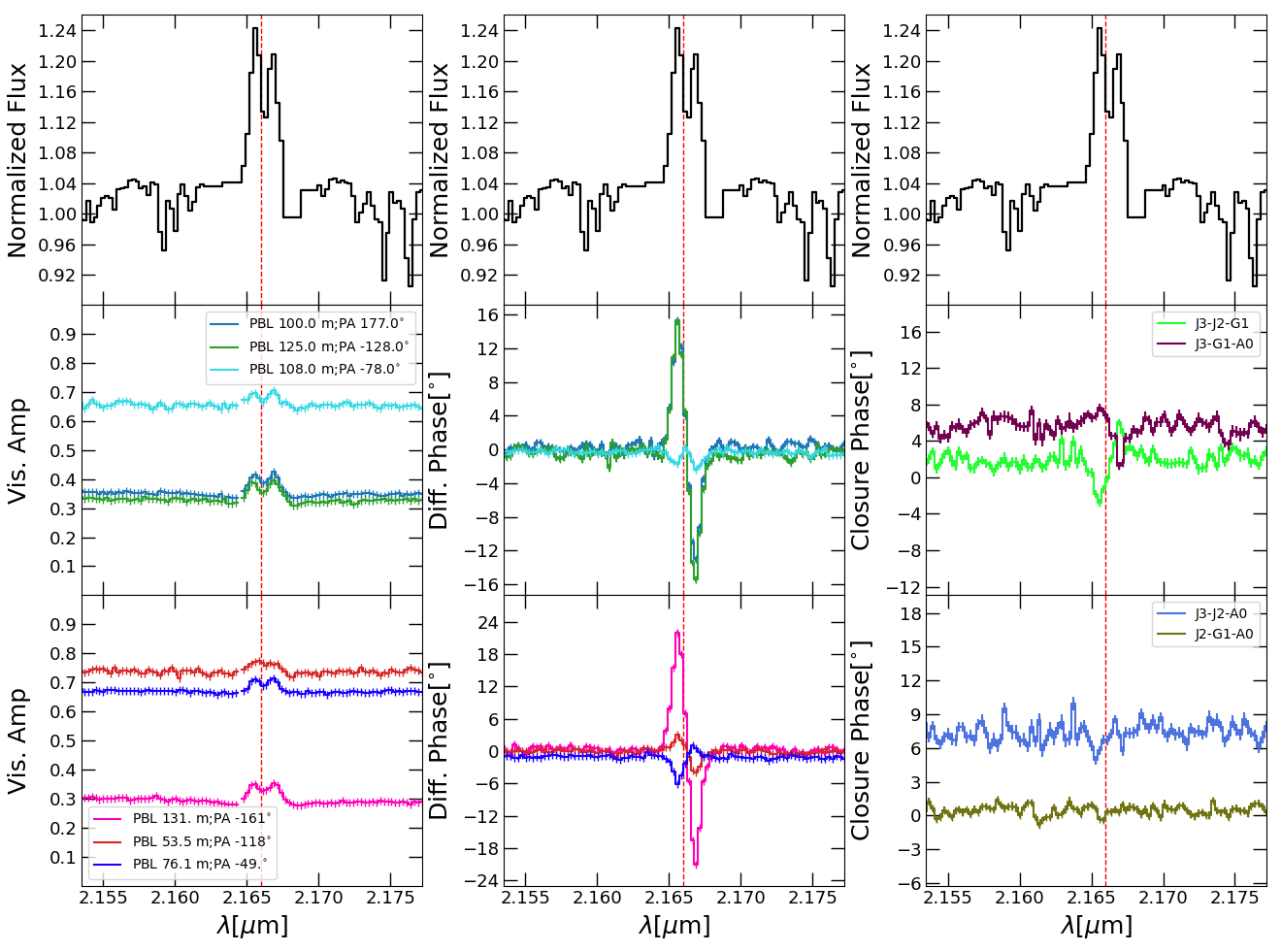}

    \caption{Same as Fig.\ B.1, but for February 10, 2021.}
\end{figure*}
\begin{figure*}[!h]
    \centering
    \includegraphics[width=\linewidth]{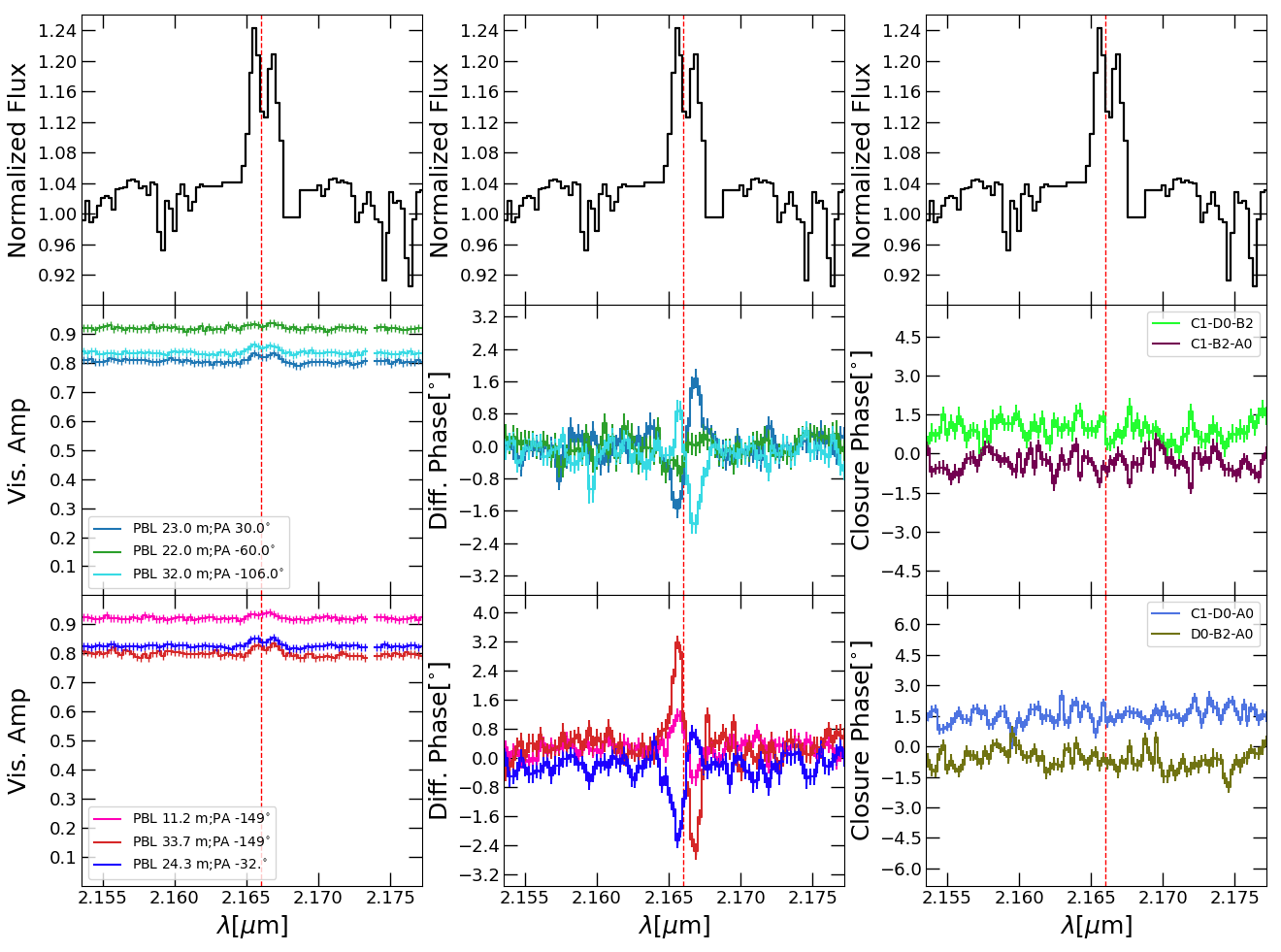}
    
    \caption{Same as Fig.\ B.1, but for February 21, 2021.}
\end{figure*}

\section{Image reconstruction parameters}
Table \ref{recparam} shows  the parameters of the image reconstructions of HD 58547.
\begin{table*}[!th]
\centering
\caption{HD\,58647 image reconstruction parameters.}
\begin{tabular}{lllllllll}
\toprule
\toprule
\multicolumn{9}{c}{Continuum}                                                                                                     \\ 
\midrule
Detector & $\lambda$ & Regularisation & Pixel scale & Grid size & $\chi^2_{\nu}$  & N & N$_{chain}$ &  Observables       \\ 
\midrule
SC         & 2.0307       & TV + l$_0$      & 0.15        & 129$\times$ 129       &  3.71   & 500    & 115     &  V2 + T3${\phi}$          \\
SC         & 2.0620     & TV + l$_0$        & 0.15        & 129$\times$ 129       & 3.01    & 500    & 115     &  V2 + T3${\phi}$           \\
SC         & 2.1235     & TV + l$_0$        & 0.15        & 129$\times$ 129       & 3.87    & 500    & 115     &  V2 + T3${\phi}$           \\
SC         & 2.1865     & TV + l$_0$        & 0.15        & 129$\times$ 129       & 2.41    & 500    & 115     &  V2 + T3${\phi}$           \\
SC         & 2.2463     & TV + l$_0$        & 0.15        & 129$\times$ 129       & 1.71    & 500    & 115     &  V2 + T3${\phi}$           \\
SC         & 2.3079     & TV + l$_0$        & 0.15        & 129$\times$ 129       & 1.69    & 500    & 115     &  V2 + T3${\phi}$           \\
SC         & 2.3689     & TV + l$_0$        & 0.15        & 129$\times$ 129       & 1.15    & 500    & 115     &  V2 + T3${\phi}$           \\
\midrule
\multicolumn{9}{c}{\brg \, Line}                                                                                                          \\
\midrule
SC         & 2.1653     & TV + l$_0$          & 0.15         & 129$\times$129       & 1.17    & 500   & 115     &  Vis$_{amp}$  + Vis$_{\phi}$ \\
SC         & 2.1656     & TV + l$_0$          & 0.15         & 129$\times$129       & 1.18    & 500   & 115     &  Vis$_{amp}$  + Vis$_{\phi}$ \\
SC         & 2.1658     & TV + l$_0$          & 0.15         & 129$\times$129       & 1.58    & 500   & 115     &  Vis$_{amp}$  + Vis$_{\phi}$ \\
SC         & 2.1661     & TV + l$_0$          & 0.15         & 129$\times$129       & 1.50    & 500   & 115     &  Vis$_{amp}$  + Vis$_{\phi}$ \\
SC         & 2.1664     & TV + l$_0$          & 0.15         & 129$\times$129       & 1.52    & 500   & 115     &  Vis$_{amp}$  + Vis$_{\phi}$ \\
SC         & 2.1666     & TV + l$_0$          & 0.15         & 129$\times$129       & 1.42    & 500   & 115     &  Vis$_{amp}$  + Vis$_{\phi}$ \\
SC         & 2.1669     & TV + l$_0$          & 0.15         & 129$\times$129       & 1.33    & 500   & 115     &  Vis$_{amp}$  + Vis$_{\phi}$ \\
\bottomrule
\end{tabular}
\label{recparam}
\end{table*}

\newpage
\section{Geometrical modelling of the K-band continuum}
\label{sect:geometrical_modelling}

We assumed that the $K$-band continuum detected by GRAVITY has two main components the stellar and circumstellar contribution. In order to take into account possible over-resolved emission, an over-resolved component (halo) has been also taken into account.  
The total complex visibility at the spatial frequencies $\left(u,v\right)$ can be written as described in \citep{Lazareff2017, Perraut2019}:\\

\begin{equation}
    V\left(u,v\right) = \frac{F_{\star}\left(\lambda_0/ \lambda \right)^{ks}V_{\star} + F_c \left(\lambda_0/ \lambda \right)^{kc} V_{c} \left(u^{\prime},v^{\prime}\right)\times {\rm e}^{-2\pi j \left(\Delta x u + \Delta y v\right)} }{\left(F_{\star}+ F_h\right)\left(\lambda_0/ \lambda \right)^{ks}+F_c\left(\lambda_0/ \lambda \right)^{kc} }
,\end{equation}
where $F_{\star},F_c , F_h$ are the different flux contributions to the total flux from the star, the circumstellar environment and the halo, respectively. The total flux is normalised at the reference wavelength $\lambda_0$. At the distance of HD\,58647 the stellar component is unresolved, and therefore $V_{\star}=1$. Finally, $k_s$ and $k_c$ are the spectral indices of the stellar and the circumstellar components defined as
\begin{equation}
    k_i = \frac{d\log{F_{\nu}}}{d\log{\nu}}
.\end{equation}
As the stellar photosphere can be approximated by a black-body, the spectral index value of the star $k_s$ is calculated from the reported surface temperature in Table\,\ref{stellarpar} and considered as a fix parameter in the fitting. In order to derive the size, inclination and PA of the circumstellar emission, a 2D Gaussian brightness distribution was assumed:

\begin{equation}
    V_{c} \left(u^{\prime},v^{\prime}\right) = \exp(-\frac{\left(\pi a \sqrt{u^{\prime 2}  + v^{\prime 2}}\right)^2}{4 \ln{2}})
\label{2dGau}
,\end{equation}
where $u^{\prime},v^{\prime}$ are the spatial frequencies corrected for the PA of the semi-major axis of the disk and its inclination, $i$,  and
\begin{equation}
\begin{split}
    &u^{\prime} = u \cos PA + v \sin PA\\
    &v^{\prime} = \left( -u \sin PA + v \cos PA \right) \cos i
\end{split}
.\end{equation}

In this model, the point-like source is centred in the phase reference point and the 2D Gaussian is displaced by $\Delta x$ and $\Delta y$ in right ascension and declination to account for possible displacements from centro-symmetry.

\begin{table}[h!]
\centering
\caption{K-band continuum model-fitting results with 1$\sigma$ error bars.}
\begin{tabular}{l c c c   }
\toprule
\toprule
Reference wavelength [$\mu m$]       & 2.18       \\
\hline
 nV$^{2}$ + nT3$\phi$                  & 4662 \\
\hline
PA [$^{\circ}$] & 13.67 $\pm$ 0.14  \\

i [$^{\circ}$] & 64.71 $\pm$ 0.12\\

a [mas] & 3.69 $\pm$ 0.02\\

$\Delta x$ [mas] & 0.200 $\pm$ 0.003\\
            

$\Delta y$ [mas] & -0.080 $\pm$ 0.003\\

$F_{\star}$ & 0.310 $\pm$ 0.001\\
$F_h$ & 0.090 $\pm$ 0.001\\

kc & -4.53 $\pm$ 0.02\\
\midrule
$\chi^2_{\nu}$ (V$^{2}$ + T3$\phi$)& 4.5 \\

\bottomrule

\end{tabular}
\label{bfpCon}
\end{table}

In order to increase the  signal-to-noise ratio (S/N), we re-binned the SC HR observables from 1634 to 7 spectral channels. The differential observable across the \brg\, line were bracketed  to keep only the continuum contribution. A Levenberg-Marquardt algorithm implemented in lmfit \citep{Newville2018} was used for the minimisation. The squared visibilities and the closure phases were fitted simultaneously at all spectral channels. The results of our best fitting model are reported in Table\,\ref{bfpCon}. A comparison between the observations and the best-fitting geometrical model is displayed in Fig.\,\ref{v2cp}. 

Our geometric model-fitting  gives a size (FWHM) of 3.69$\pm$0.02\,mas (i.\,e. 1.114$\pm$0.006\,au at a distance of 302\,pc), and an inclination of $\mathrm{i}$=64.71\degr$\pm$0.12\degr, with a disk major-axis PA (N to E) of 13.67\degr$\pm$0.14$^{\circ}$. We note that the error bars on the derived parameters resulting from the model fitting represent only the formal errors arising from the fitting procedures, and do not account for other sources of error.

The derived size is in agreement with \cite{Perraut2019}. As, expected our continuum $K$-band size (FWHM$\sim$ 3.7\,mas) is larger than the $H$-band size (FWHM$\sim$3\,mas), which is consistent with a cooler emission in the $K$-band  than in the $H$-band.
In addition, our modelling suggests the presence of more extended emission as traced by the over-resolved component in our model. The flux ratio associated with this component is of $\sim$9\% of the total flux. Again this is in agreement with previous H-band results \citep{Lazareff2017}, although the H-band over-resolved component has a much smaller contribution to the total flux than the one reported here (flux ratio of 2\% vs the 9\% observed in our data). The origin of this over-resolved component is still under debate and it might be attribute to the presence of scattered light \citep[e.g.][]{Pinte2008, Benisty2010}. In order to obtain more information about this component a better u-v coverage as well as additional information from short baselines are needed in order to constrain its nature and morphology.

Finally, it should be pointed out that although our geometrical model is roughly reproducing the general trend of the visibilities and closure phase signatures with spatial frequency, it is poorly recovering their exact values at some spatial frequencies, especially at the shortest baselines (see Fig.\,\ref{v2cp}). This is reflected by the observed residuals  ($\chi^2_{\nu}\,\sim 4.5$) and can be explained by the fact that our model is too simple to reproduce the complexity of the structure at the largest spatial scales probed by those baselines. To fully investigate the nature of the departures from centro-symmetry, image reconstruction is needed (see Sect.\,\ref{sec4}).

\begin{figure*}[ht!]
\centering
\includegraphics[width=\linewidth]{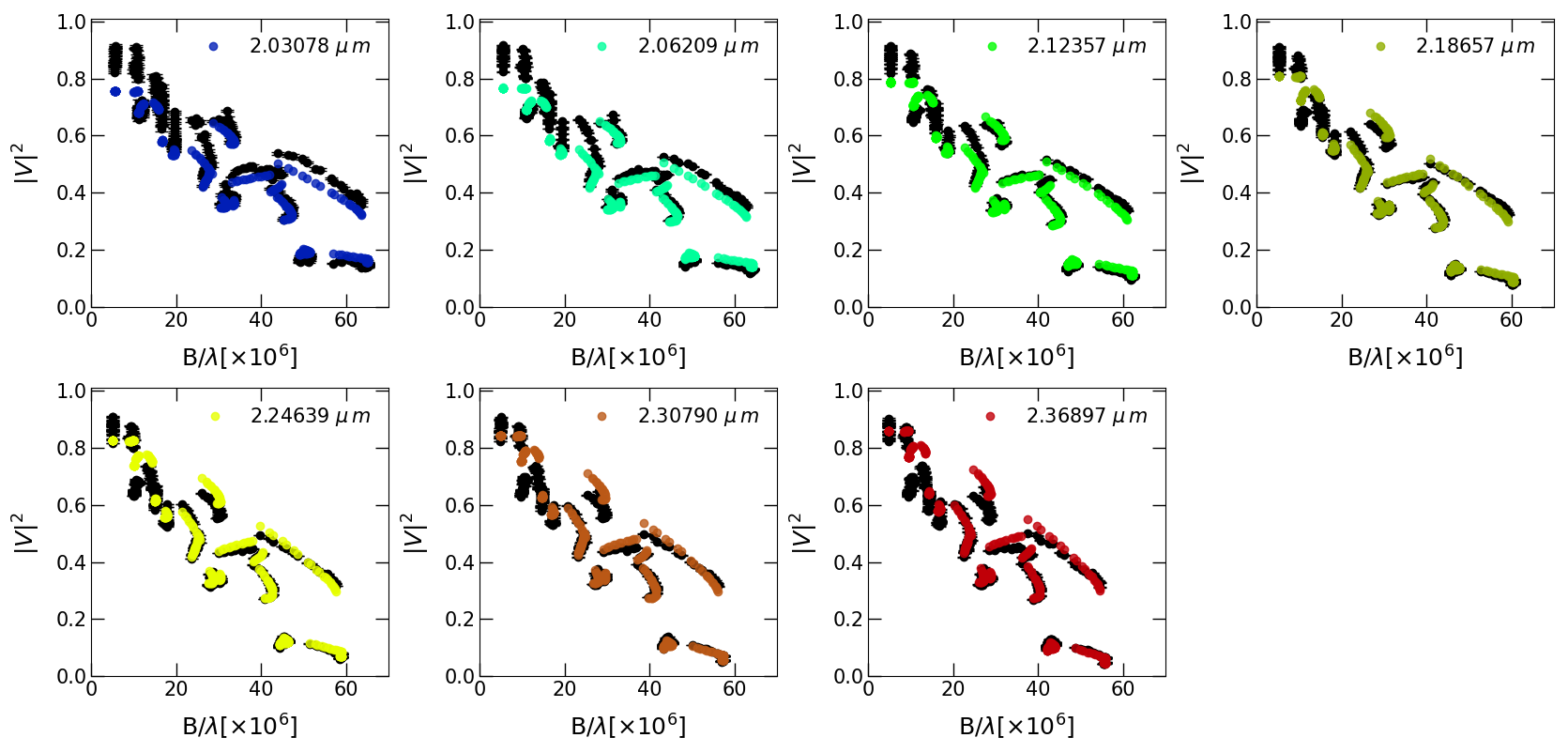}
\includegraphics[width=\linewidth]{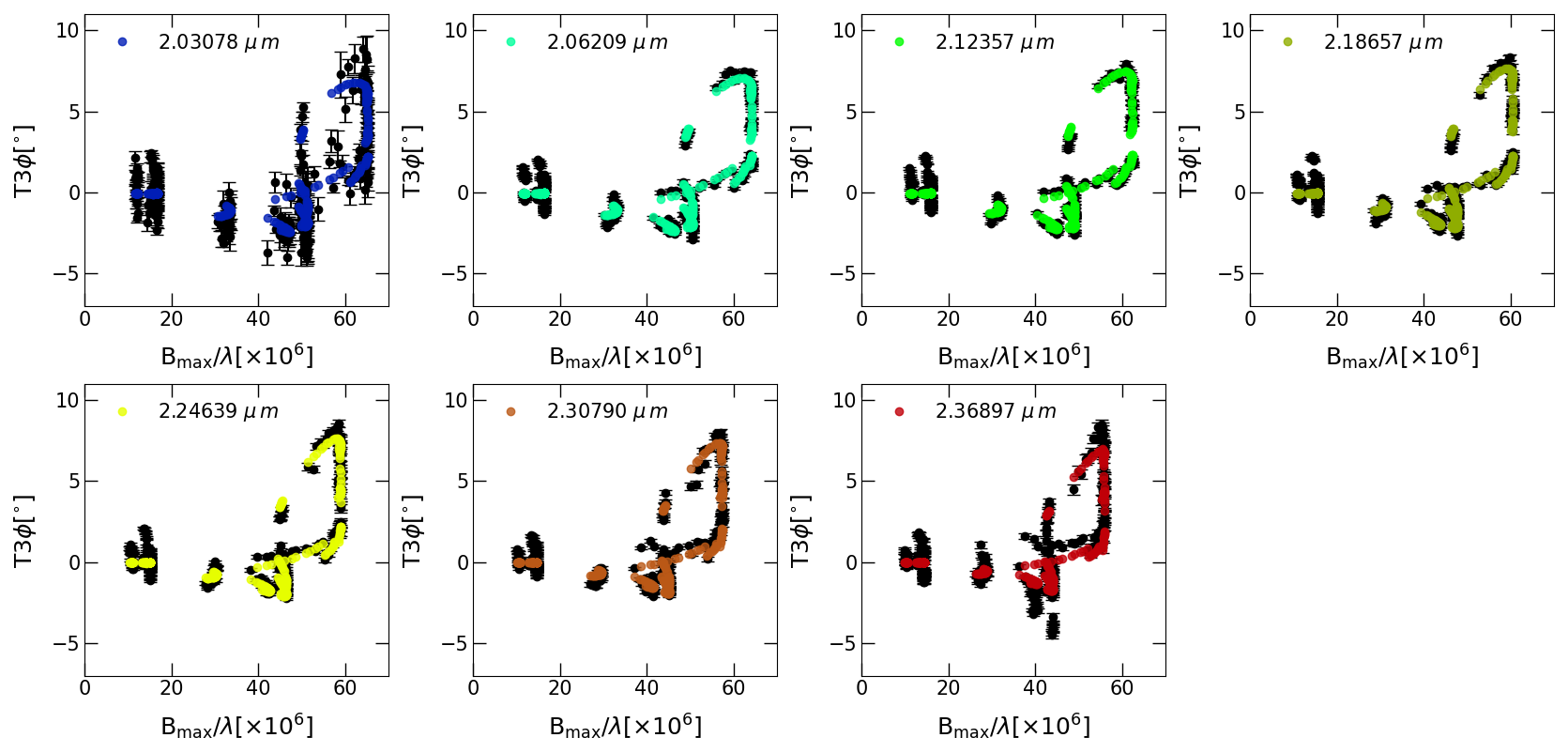}
\caption{Comparison between the observed squared visibilities ($\left | V\right |^2 $) and closure phase signatures (T3$\phi$)  of HD\,58647 (black dots and error bars) with the results from a 2D Gaussian brightness distribution (coloured dots).  }
\label{v2cp}
\end{figure*}

\section{\brg \, line continuum-corrected visibilities and differential phases}
\label{sect:continuum-subtracted}

The HR  differential visibilities show an increase across the $\mathrm{Br}\gamma$ line at all baselines indicating that the line emitting region is more compact than the continuum. However, the observed differential observables are contaminated by the continuum emission and thus this contribution needs to be subtracted.
In order to estimate the size of the $\mathrm{Br}\gamma$ line emitting region we took the contribution from  the continuum emission into account  and calculated the continuum-corrected visibilities (see \cite{Weigelt2007} for more details):

\begin{equation}
    \left|F_{\mathrm{line}} V_{\mathrm{line}}\right|^2 = \left|F_{\mathrm{tot}} V_{\mathrm{tot}}\right|^2 + \left|F_c V_c\right|^2 - 2 \, F_{\mathrm{tot}} V_{\mathrm{tot}}\, F_c V_c \cos{\phi}
\label{vline}    
,\end{equation}
where $F_{\mathrm{tot}}$ is the total observed flux, $V_{\mathrm{tot}}$ the observed visibility, $F_{\mathrm{line}}$ the flux line, $V_c$ and $F_c$ the continuum visibility and the continuum flux inside the $\mathrm{Br}\gamma$ line,  and $\phi$ describes the observed differential phases.

We used a polynomial interpolation to calculate the flux of the continuum across the $\mathrm{Br}\gamma$ line assuming that the flux of the continuum outside the  line is  equal to the level of the continuum inside the line. The visibility amplitude of the continuum across the line was also estimated from a polynomial interpolation. The continuum channels nearby the $\mathrm{Br}\gamma$ line were used to perform the interpolation.

We calculated the line visibility for seven spectral channels where the flux of the line is 10\% higher than that of the continuum flux and the errors are estimated by propagating Eq. \ref{vline}. We used the 1$\sigma$ errors as estimated by the data reduction software for the visibility amplitudes and phases. We assumed that the \brg\ line emitting region has the following components: an unresolved component with a flux ratio described by $f_s$, and over-resolved component $f_h$, and finally a 2D Gaussian (similar to Eq.\,\ref{2dGau}) with visibility V associated with a flux ratio $f_c$ such as  $f_s + f_h + f_c = 1$ describing the resolved component. The total visibility is then given by

\begin{equation}
    V_{amp}\left(u,v\right) = f_s + \left(1-f_s - f_h \right) V \left(u^{\prime},v^{\prime}\right)
.\end{equation}

We chose to fit the visibilities across the line  at blue { (channels at $\sim$-72\,\kms; 2.1655)}, red { (channels at $\sim$0\,\kms;  2.1661 \,\um)}, and zero { (channels at $\sim$109\,\kms ; 2.1668 \,\um)} velocities.
The model-fitting parameters are listed in Table\,\ref{bfpLine}. The blue- and redshifted line emitting regions have a similar sizes, inclinations and PA.  
The results of our fitting show that about 50\% of the flux is coming from an unresolved component with negligible contribution from an over-resolved component. However, the results shown in Table\,\ref{bfpLine} should be taken with caution. Unlike  the K-band continuum model-fitting, our simple 2D Gaussian brightness distribution does not reproduce the complex geometry of the line emitting region and only the visibility amplitude is modelled (see Fig.\,\ref{vampfitting} for a comparison between the model-fitting results and the observations). In order to further probe the structure of the system image reconstruction is needed (see Sect.\,\ref{sec5}).

\begin{figure*}[ht!]
\centering
\includegraphics[width=\linewidth]{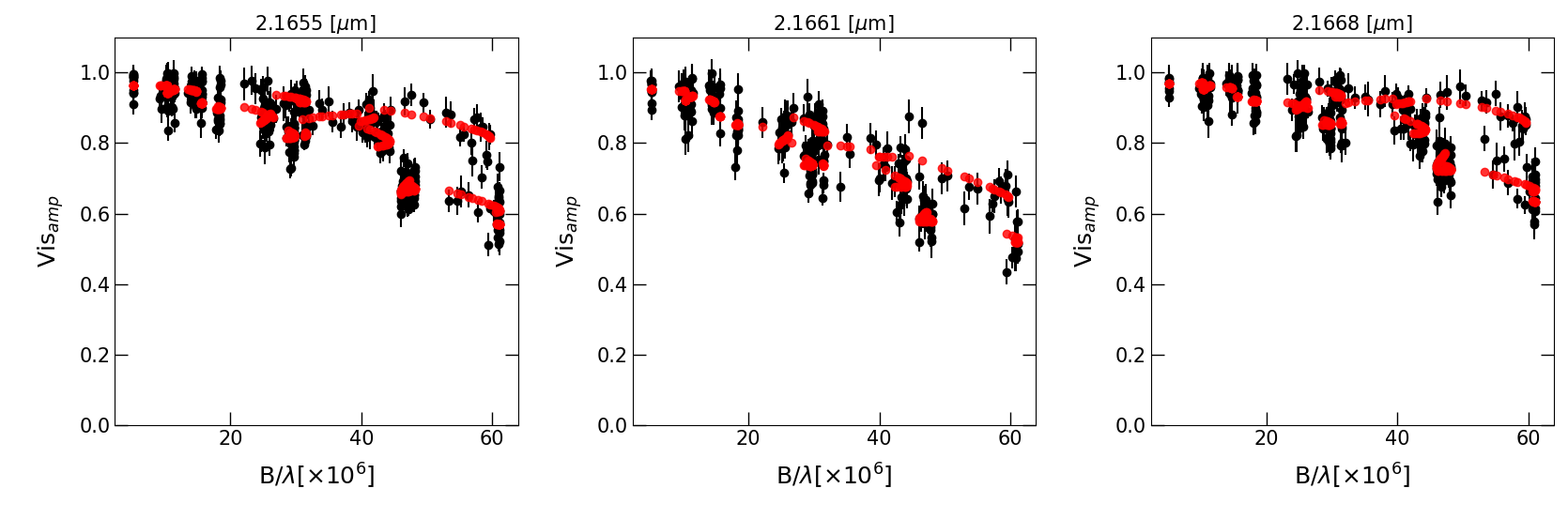}
\caption{\brg\ line model-fitting results (red dots) as a function of the spatial frequency at three different spectral channels. For comparison, the continuum-corrected \brg\ line visibilities are overplotted in black.}
\label{vampfitting}
\end{figure*}

\begin{table}[h]
\caption{\brg\ line model-fitting results.}
\centering
\begin{tabular}{l c c c c  }
\toprule
\toprule
Wavelength [$\mu m$]       & 2.1655  & 2.1661  & 2.1668     \\
\hline
PA [$^{\circ}$] &  15$^{+0.83}_{-0.82}$ & 17.35$^{+1.35}_{-1.34}$ & 19.31$^{+1.05}_{-1.05}$  \\

i [$^{\circ}$] &  67$^{+0.88}_{-0.88}$  & 55.27$^{+0.94}_{-0.95}$ & 68.95$^{+1.41}_{-1.35}$\\

a [mas] & 2.6$^{+0.08}_{-0.08}$ & 3.23$^{+0.08}_{-0.08}$ & 2.32$^{+0.11}_{-0.11}$\\

fs [\%] &  51.32$^{+1.43}_{-1.57}$ & 50.03$^{+1}_{-1}$ & 55.43$^{+2.02}_{-2.36}$\\
fh [\%] &  3.1$^{+0.3}_{-0.3}$ & 3.77$^{+0.4}_{-0.4}$ & 2.64$^{+0.34}_{-0.34}$\\    
$\chi^2_{\nu}$ & 0.94    &  0.92 & 0.91\\

\bottomrule
\end{tabular}
\label{bfpLine}

\end{table}

\section{Continuum-corrected differential phases}
\label{diff-phases}

The observed differential phases (see Fig.\,\ref{diffobs} bottom panel) show deviation from zero and allow us to study the displacement of the photocentre of the emission across the $\mathrm{Br}\gamma$ line with respect to the continuum. In order to analyse any displacement of the photocentre of the emission, the differential phase needs to be corrected for the continuum emission:
\begin{equation}
    \sin{\phi_\mathrm{line}} = \sin{\phi}\frac{\left|F_{\mathrm{tot}} V_{\mathrm{tot}}\right|}{\left|F_\mathrm{line} V_\mathrm{line}\right|}
,\end{equation}
where $\phi_\mathrm{line}$ is the continuum-corrected differential phase and $\phi$ is the observed differential phase \citep[see][for more details]{Weigelt2007}.

By combining all the continuum-corrected differential phases, we can derive a 2D displacement of the photocentre \citep[see][for additional details]{Lachaume2003, LeBouquin2009}:

\begin{equation}
    \vb{P} = -\frac{\phi_{line}^i}{2\pi} \cdot \frac{\lambda}{\vb{B_i}}
.\end{equation}

The astrometric displacements vectors  projected on the sky are shown in Fig.\,\ref{2ddisp}. The figure shows all displacements lying along a straight line, with a clear measurement of the displacement between the  red-  and blueshifted velocities. By a linear fit of the displacements, a PA  of the line emitting region of {$\sim$15\degr$\pm$5\degr }
is estimated. This value is in agreement with that derived from the geometric model of the continuum disk. 

\begin{figure}[th!]
\centering
\includegraphics[width=\columnwidth]{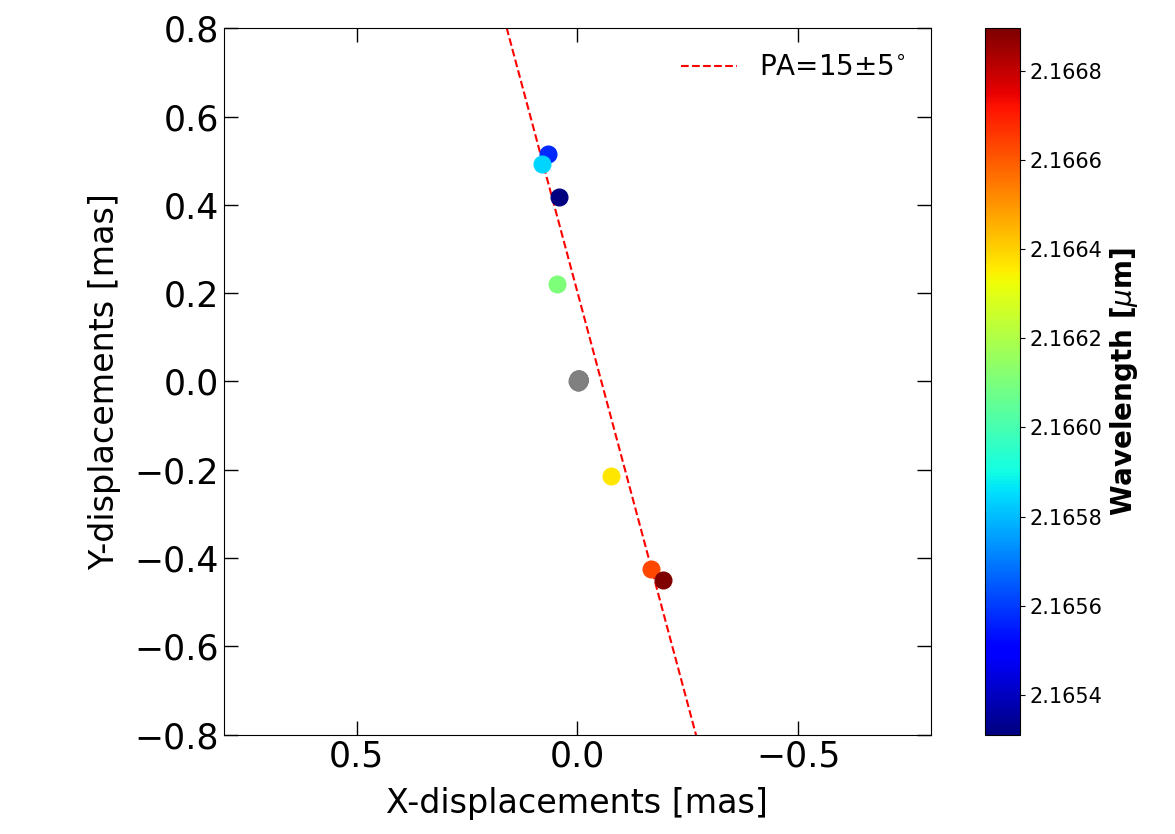}
\caption{Astrometric displacements obtained from the \brg\ line continuum-corrected differential phases.}
\label{2ddisp}
\end{figure}

\begin{figure*}[!h]

\includegraphics[width=\textwidth]{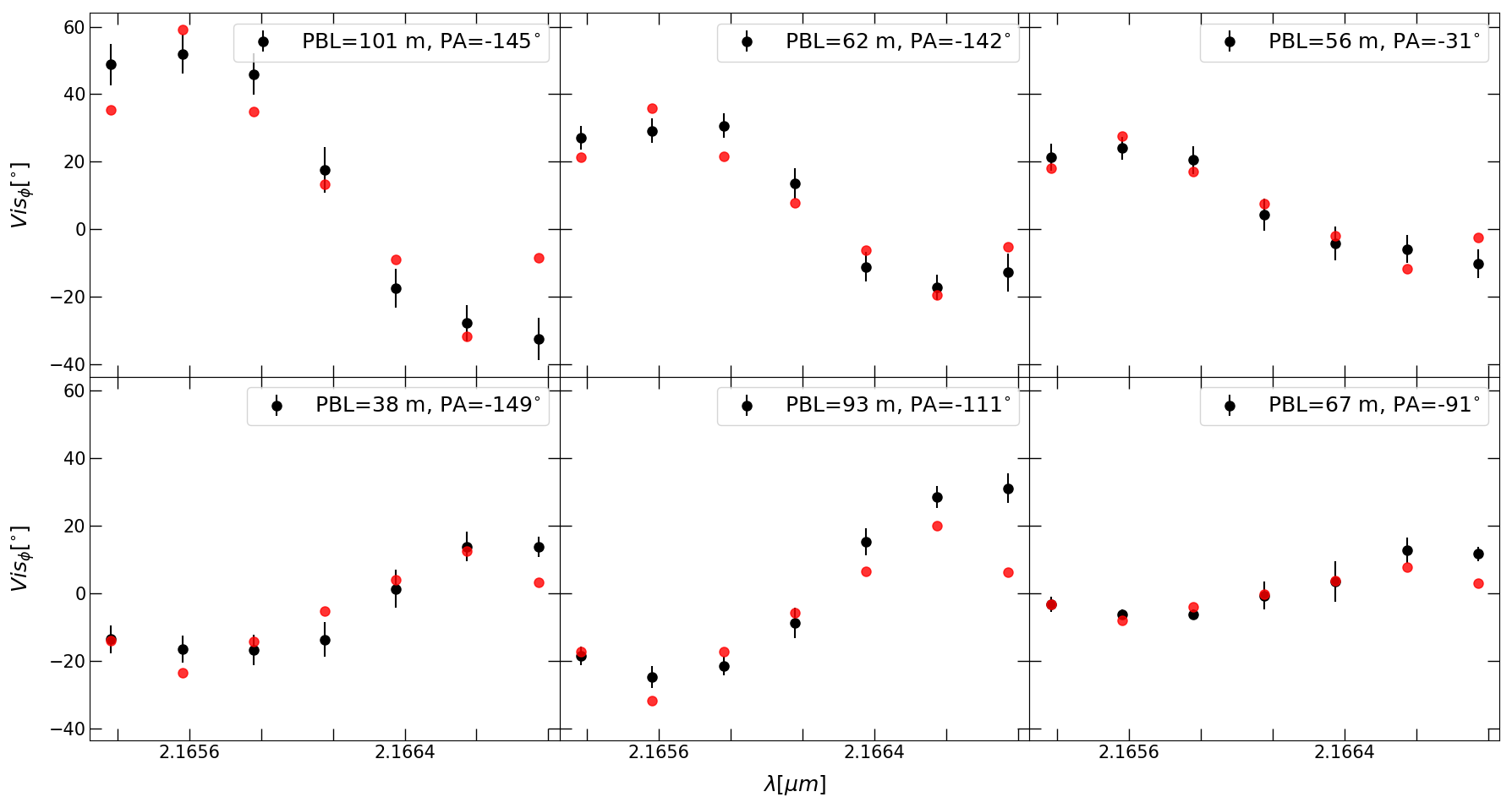}
\caption{Example of the synthetic differential phases obtained from the \brg\ Keplerian disk model (red dots). The observed continuum-corrected \brg\ line  phases are shown in black.}
\label{ccdiffph_old}
\end{figure*}

\end{appendix}

\end{document}